\documentclass[a4paper,fleqn,usenatbib,useAMS]{mnras}
\usepackage{graphicx}	
\usepackage{amsmath}	
\usepackage{amssymb}	
\usepackage{multicol}        
\usepackage{bm}		
\usepackage{pdflscape}	
\usepackage{morefloats}
\usepackage{longtable}
\usepackage{geometry} 
\usepackage{booktabs,longtable}
 
\newcommand{\hii}{\,H\,{\small II}}
\newcommand{\hi}{\,H\,{\small I}}
 
\usepackage[T1]{fontenc}
\usepackage{ae,aecompl}
\title[North American and Pelican cloud complexes]{Testing the star formation scaling relations in the clumps of the North American and Pelican cloud complexes }

\author[Swagat R Das]{Swagat R. Das$^{1}$\thanks{Contact e-mail: \href{dasswagat77@gmail.com}{dasswagat77@gmail.com}.}
, {Jessy Jose}$^{1}$, {Manash R Samal}$^{2}$, {Shaobo Zhang}$^{3}$, \\
\newauthor and {Neelam Panwar}$^{4}$ 
\\ \\
$^{1}$Indian Institute of Science Education and Research (IISER) Tirupati, Rami Reddy Nagar, Karakambadi Road, \\ 
Mangalam (P.O.), Tirupati 517507, India \\
$^{2}$Physical Research Laboratory,Ahmedabad, Gujrat, India \\
$^{3}$Purple Mountain Observatory, \& Key Laboratory for Radio Astronomy, Chinese Academy of Sciences, Nanjing 210023, China \\
$^{4}$Aryabhatta Research Institute of observational sciencES (ARIES), Manora Peak, Naini Tal 263129, India \\
}

\begin{document}
\label{firstpage}
\pagerange{\pageref{firstpage}--\pageref{lastpage}}
\maketitle

\begin{abstract}
The processes which regulate the star-formation within molecular clouds are still not well understood. Various star-formation scaling relations have been proposed to explain this issue by formulating a relation between star-formation rate surface density ($\rm \Sigma_{SFR}$) and the underlying gas surface density ($\rm \Sigma_{gas}$). In this work, we test various star formation scaling relations, such as Kennicutt-Schmidt relation, volumetric star-formation relation, orbital time model, crossing time model, and multi free-fall time scale model towards the North American and Pelican Nebulae complexes and in cold clumps associated with them. Measuring stellar-mass from young stellar objects and gaseous mass from CO measurements, we estimated mean $\rm \Sigma_{SFR}$, star formation rate per free-fall time, and star formation efficiency (SFE) for clumps to be 1.5~$\rm M_{\odot} yr^{-1} kpc^{-2}$, 0.009, 2.0\%, respectively, while for the entire NAN complex the values are 0.6~$\rm M_{\odot} yr^{-1} kpc^{-2}$, 0.0003, and 1.6\%, respectively. For clumps, we notice that the observed properties are in line with the correlation obtained between $\rm \Sigma_{SFR}$ and $\rm \Sigma_{gas}$, and between $\rm \Sigma_{SFR}$ and $\rm \Sigma_{gas}$ per free-fall time and orbital time for Galactic clouds. At the same time, we do not observe any correlation with $\rm \Sigma_{gas}$ per crossing time and multi free-fall time. Even though we see correlations in former cases, however, all models agree with each other within a factor of 0.5~dex, and discriminating between these models is not possible due to the current uncertainties in the input observables. We also test the variation of $\rm \Sigma_{SFR}$ versus the dense gas, but due to low statistics, a weak correlation is seen in our analysis.

\end{abstract}



\section{Introduction} \label{intro}
Molecular clouds are the dense regions of the interstellar medium (ISM), which supplies all the raw materials for the formation of stars. Star-formation is a multi-stage process in which the dense part of molecular clouds undergo gravitational collapse and set footprints for the formation of new stars. Though tremendous progress has been achieved in recent decades, the understanding of complete episodes of star-formation is still under debate. This is a fundamental issue in the field of astrophysics.

The process which regulates the conversion of gas into stars is still least understood. \citet{1959ApJ...129..243S} was the first who proposed a SFR-gas relation (popularly known as ``Schmidt Relation") in the form of a power law, in which the star-formation rate (SFR) surface density is proportional to the square of volume density. Later star-formation properties of many spiral and starbursts galaxies were studied by \citet{1998ApJ...498..541K} using both  atomic (\hi) and molecular ($\rm H_2$) gas for the estimation of $\rm \Sigma_{gas}$ in the entire galaxies. He proposed a power law between the star-formation rate density ($\rm \Sigma_{SFR}$) and gas surface density ($\rm \Sigma_{gas}$) in the form of $\rm \Sigma_{SFR}$ $\rm \propto$ $\rm \Sigma_{gas}^N$. This scaling relation is called as the ``Kennicutt--Schmidt relation", where N = 1.4. 
Using sensitive observations with sub-kpc resolution, \citet{2008AJ....136.2846B} studied the atomic and molecular gas in 18 nearby galaxies and obtained a linear relation between $\rm \Sigma_{SFR}$ and molecular gas density, for values above 10~$\rm M_{\odot}~pc^{-2}$. According to this study, the value of N is $1.0~\pm~0.02$. \citet{2015ApJ...805...31L} have analyzed 181 galaxies, out of which 115 are normal spiral galaxies, and 66 are IR galaxies. These authors reported a tighter correlation between $\rm \Sigma_{gas}$ (traced by HCN) and $\rm \Sigma_{SFR}$, with N = 1.01$\pm$0.02. In a recent study, \cite{2019ApJ...872...16D} examined 169 spiral and 138 dwarf galaxies. These authors found that spiral galaxies define a tight relation between $\rm \Sigma_{SFR}$ and $\rm \Sigma_{gas}$ (estimated from both atomic \hi~ and molecular $\rm H_2$) with value of N = 1.41$\pm$0.07. According to this study, the $\rm \Sigma_{SFR}$ are only weakly correlated with \hi~ surface densities, but exhibit a
stronger and roughly linear correlation with $\rm H_2$ surface densities. The value of N was found generally to be in between $1 - 2$ from observations of several galaxies carried out in last two decades \citep{{2010A&A...521A..48S},{2003MNRAS.346.1215B},{2001ApJ...555..301M},{2004ApJ...602..723H},{2005PASJ...57..733K},{2007A&A...461..143S}}.

Most of these above results are based on studies of either resolved observations of galaxies or the complete galaxies. However, many studies have been carried out to test the relation of $\rm \Sigma_{SFR}$ and $\rm \Sigma_{gas}$ in Milky Way clouds and compare them with the extragalactic relations. \citet{2005ApJ...635L.173W} have observed dense cores using HCN, both in external galaxies and in our Galaxy, and found a tight SFR-gas relationship, which was earlier observed by \citet{2004ApJ...606..271G} towards a sample of IR galaxies. 
\citet{2009ApJS..181..321E} have examined several regions from {\it Spitzer} c2d survey and found that the SFR-gas relation for the Galactic clouds lie above the relations of \citet{1998ApJ...498..541K} and \citet{2008AJ....136.2846B}, but lie slightly above the relation of \citet{2005ApJ...635L.173W}. \citet{2009ApJS..181..321E} found that the $\rm \Sigma_{SFR}$ of Galactic clouds are higher by a factor $\sim20$ compared to the values obtained using the relation of \citet{1998ApJ...498..541K}. \citet{2010ApJ...723.1019H} examined seven c2d regions and 13 regions from the Gould Belt (GB) survey and extended the comparisons of Galactic SFR-gas relations with the extragalactic relations. They also found a similar conclusion as obtained by \citet{2009ApJS..181..321E}.  \citet{2010ApJ...723.1019H} reported that their $\rm \Sigma_{SFR}$ is higher by a factor of $\sim30$ compared to the values obtained using relation of \citet{1998ApJ...498..541K}. This discrepancy is mainly due to the fact that the Kennicutt--Schmidt relation \citep{1998ApJ...498..541K} is derived over large regions such as galaxies, where the diffused clouds within the galaxies, have no role in active star-formation. \citet{2010ApJ...723.1019H} also derived a relation similar to that of \citet{2005ApJ...635L.173W}.

The simple Kennicutt--Schmidt relation shows a large scatter between $\rm \Sigma_{SFR}$ and $\rm \Sigma_{gas}$. Many discussions found in literature, for the search of a better correlation between $\rm \Sigma_{SFR}$ and $\rm \Sigma_{gas}$. 
\citet{2010ApJ...723.1019H} have suggested that above a certain threshold gas surface density, the star formation rate and the cloud mass are better correlated. From their studies, $\rm A_V \sim 8~mag $ found to be the threshold level, which was also confirmed independently by \citet{2010ApJ...724..687L}. However, later on, some studies have questioned the existence of a density threshold in molecular clouds \citep{{2013ApJ...773...48B},{2014MNRAS.444.2396C}}. To reduce the scatter between the SFR-gas variance observed in Kennicutt--Schmidt relation, \citet{2012ApJ...745...69K} have proposed the volumetric star-formation relation which is a  congruity between $\rm \Sigma_{SFR}$ and $\rm \Sigma_{gas}/t_{ff}$, where $\rm t_{ff}$ is the free-fall time scale. According to these authors, the volumetric star-formation relation is able to reduce the high scatter seen in SFR-gas variance obtained by \citet{1998ApJ...498..541K}. This relation has been examined extensively by \citet{2012ApJ...745...69K} and \citet{2013MNRAS.436.3167F}. In these works, the free-fall time scale is estimated over the average density of the cloud. \citet{2014ApJ...782..114E} tested this relation and found that the free-fall model did not work well for nearby clouds.
It is suggested that due to the clumpy nature of molecular clouds, the typical free-fall time scale of star-formation should not be the average time scale of the cloud. Rather it is the density-dependent time scale of the substructures which collapse to form stars \citep{2015ApJ...806L..36S}. This is called the multi free-fall time concept \citep{{2011ApJ...743L..29H},{2013ApJ...770..150H},{2014ApJ...796...75C}} in which the variation of $\rm \Sigma_{SFR}$ with $\rm \Sigma_{gas}/t_{multi-ff}$ is examined. A relation of $\rm \Sigma_{SFR}$ with $\rm \Sigma_{gas}$ evaluated over the orbital time scale is suggested by \citet{1998ApJ...498..541K}. In this case, it is proposed that the star-formation is affected by galactic spiral arms and bars. These large scale radial processes affect the star-formation occurring from molecular gas over an orbital time scale \citep{{1989ApJ...339..700W},{1998ApJ...498..541K}}. Star-formation evolved over several cloud crossing time scales have been studied in detail by \citet{2000ApJ...530..277E}. In this study, it was suggested that the star-formation process is rapid in dense regions, and it finishes in several cloud crossing time scales.

In summary, there exists a handful of star formation scaling relations to explain the star formation processes in molecular clouds. The North American and Pelican Nebulae complexes are young, massive, and {is a} nearby Galactic star-forming region. For this region, deep multi-wavelength data is available to make a complete census of young stellar objects (YSOs). This motivates us to analyze the various star-formation relations in dust clumps associated with North American and Pelican Nebulae complexes. We have explained various star-formation scaling relations in more detail in later sections, while we test them towards our region. 
This work has been presented in the following way. In Section 2, we discuss details about the target. In Section 3, we discuss the details of the data sets used in this study. In Section 4, we discuss various analysis and results, and section 5 summarizes the results.

\section{Details / characteristics  of the target} \label{NAN_detail}
\begin{figure*}
\centering
\includegraphics[scale=0.5]{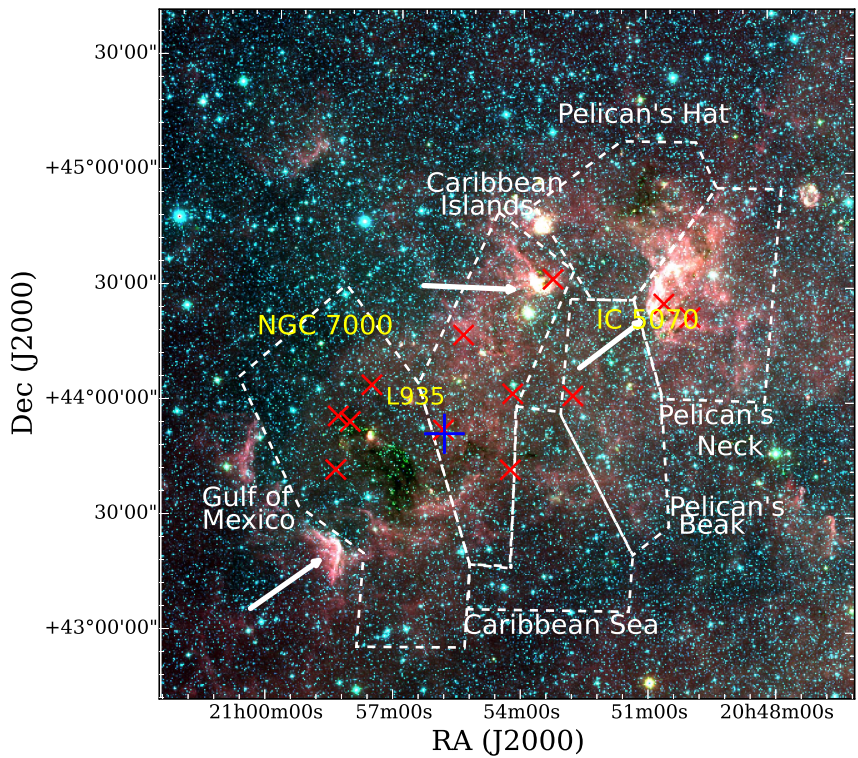}
\caption{
Color-composite image of the NAN complex made using WISE 12~$\rm \mu m$ (red), 4.6~$\rm \mu m$ (green), and 3.6~$\rm \mu m$ (blue) images. This image shows the complete NAN complex along with the locations of NGC 7000 and IC 5070. Warm dust emission is seen at 12~$\rm \mu m$, and the other two bands show the stellar population. The dashed line shows the outline of the six sub-regions within the complex identified by \citet{2014AJ....147...46Z}. Names of all the regions are also given. Blue plus sign (``+'') marks the location of ionizing star \citep{2005A&A...430..541C}. The reddened O-type stars \citep{2008BaltA..17..143S} are shown as red cross (``$\times$'') marks. The white arrows point towards the locations of the Bright-rimmed clouds within the complex.
 }
\label{wise_co}
\end{figure*}

Our analysis covers the region associated with North American Nebula (NGC 7000) and the Pelican Nebula (IC 5070). Both the regions are part of the large \hii\ region called W80 \citep{{1955ApJ...121..611M},{1958BAN....14..215W}}. We call the whole region covered by both the nebulae as ``NAN'' complex \citep{{2011ApJS..193...25R},{2014AJ....147...46Z}}. This is a nearby molecular cloud complex and associated with many massive stars. Distance measurements to this complex vary in a range of 500~pc -- 1~kpc \citep{{1968ZA.....68..368W},{1983A&A...124..116W},{1980A&A....92L...9N},{1993BaltA...2..171S},{2002BaltA..11..205L},{2006BaltA..15..483L}}. A recent estimate of distance to the complex is $\rm 800\pm40~pc$, using the GAIA DR2 data \citep{2020A&A...633A..51Z}. In our analysis we adopt this distance estimation.

NAN complex is a well studied star-forming region at near-infrared (NIR), mid-infrared (MIR), and also in millimeter (mm) and sub-mm wavebands of molecular line data. Using the 2MASS photometry, \citet{2005A&A...430..541C} have identified an O5V star (2MASS J205551.25~+~435224.6) as the ionizing source of the NAN complex. A few more reddened O-type stars are also identified by \citet{2008BaltA..17..143S}, which might play a role in ionizing the hydrogen gas in the NAN region. Using MIR data \citet{2009ApJ...697..787G} and \citet{2011ApJS..193...25R} have identified a total of 2082 YSOs in the complex. These YSOs are identified from the IRAC and MIPS observations of {\it Spitzer} Space Telescope, covering a region of $\rm \sim 7~deg^2$ centred at $\rm \alpha = 20:54:18.0$ and $\rm \delta = +44:05:00.0$. From the analysis of molecular line data \citep{{1980ApJ...239..121B},{1994ApJS...95..419D}} and NIR extinction \citep{2002AJ....123.2559C}, it has been found that a large amount of molecular gas is located toward the Dark Nebula LDN 935 \citep{1962ApJS....7....1L}, which bisects the North American and Pelican nebulae. The latest analysis of NAN complex using $\rm ^{12}CO, ^{13}CO$, and $\rm C^{18}O$ molecular line data by \citet{2014AJ....147...46Z} have identified filamentary structures and investigated the star-formation properties within them. These authors have identified several dense clouds of surface density over $\rm 500~M_{\odot}~pc^{-2}$. The mean $\rm H_2$ column densities are 5.8, 3.4, and $\rm 11.9 \times 10^{21}~cm^{-2}$ for $\rm ^{12}CO,~^{13}CO,~ and~C^{18}O$, respectively and total mass of the complex is estimated to be $\rm 5.4 \times 10^4,~2.0\times 10^4,~ and~6.1\times 10^3~M_\odot$ from $\rm ^{12}CO,~^{13}CO,~ and~C^{18}O$, respectively. This indicates the different forms of gas traced by the molecules. The mass traced by the $\rm ^{13}CO$ and $\rm C^{18}O$ lines are $\sim36\%$ and $\sim11\%$ of the mass traced by the $\rm ^{12}CO$. In their analysis, \citet{2014AJ....147...46Z} have identified 611 small scale dense cores from the $\rm ^{13}CO$ map. However, in our analysis, we are interested in large scale molecular cloud clumps and probe the star-formation properties within them. Using GAIA astrometry, the clustering and kinematics of the YSOs and associated molecular gas have been analyzed by \citet{2020arXiv200608622K}. This work has focused only on a small area around the star-forming complex consisting of some clusters. The latest spectroscopic analysis on the NAN complex by \citet{2020arXiv200911995F} suggests a sequential history of star formation in the NAN region.

In this work, we aim to probe the SFR and star formation efficiency (SFE) of dense clouds located in the NAN complex. These are two important parameters to study the star-formation activity of a star-forming complex. SFR is defined as the amount of mass converted into stars per unit time, whereas SFE is defined as the ratio of the stellar mass of star-forming region to the total stellar and gas mass of the parental cloud. Using molecular line and photometric data in deep NIR and MIR bands, we analyze the SFR and SFE of the entire NAN complex. In Figure \ref{wise_co}, we show the whole field of view of the NAN complex in MIR bands. Based on the spatial distribution of molecules, \citet{2014AJ....147...46Z} have visually identified the boundaries of the six star-forming regions identified in \citet{2011ApJS..193...25R} and their respective names are marked in Figure \ref{wise_co}, along with their boundaries. The WISE\footnote{This publication makes use of data products from the Wide-field Infrared Survey Explorer, which is a joint project of the University of California, Los Angeles, and the Jet Propulsion Laboratory/California Institute of Technology, funded by the National Aeronautics and Space Administration.} color composite image shows the presence of warm dust emission traced by the 12~$\rm \mu m$ band. The warm dust emission appears to be distributed along the border of two broken bubbles. Bright-rimmed clouds (BRCs) are located on the border of the bubbles \citep{2002AJ....123.2597O} (see Figure \ref{wise_co}). The lower bubble is located in the region of the Gulf of Mexico, and the upper bubble is distributed within regions of the Caribbean Islands, Pelican's Neck, and Pelican's Hat. A dark filament is seen within the Gulf of Mexico, which could be due to the early stage of the cloud. Four O-type stars are located within the Gulf of Mexico. The complete NAN complex is active in star formation, which is clear from the presence of various outflows, $\rm H_2$ jets, HH objects and $\rm H\alpha$ emission line stars and bright rimmed clouds \citep{{2003AJ....126..893B},{2005A&A...430..541C},{2011A&A...528A.125A},{2014AJ....148..120B}}.

\section{Data used in this study}
\subsection{{\it Spitzer MIR data}}
We use the MIR photometric data obtained from {\it Spitzer} Space Telescope \citep{2004ApJS..154....1W} with Infrared Array Camera (IRAC; \citealt{2007astro.ph..3037W}). The region towards the NAN complex studied in four IRAC bands by \citet{2009ApJ...697..787G} and in three bands of Multiband Imaging Photometer for {\it Spitzer} (MIPS; \citealt{2004ApJS..154...25R}) by \citet{2011ApJS..193...25R}. In our analysis, we obtained the photometric data in 3.6 and 4.5~$\rm \mu m$ IRAC bands used in \citet{2009ApJ...697..787G}  and \citet{2011ApJS..193...25R} (private communication). To ensure all sources are of good photometric quality, we use only those sources with error less than 0.2~mag. The IRAC and deep JHK photometry data are used to identify and classify the extra YSOs and to study the star formation properties in the NAN complex.

\subsection{NIR data from UKIDSS}
In this work, we retrieve NIR photometric data in J, H, and K bands from the UKIDSS \citep{2007MNRAS.379.1599L} Galactic Plane Survey (GPS; \citealt{2008MNRAS.391..136L}). This survey was carried out with the  UKIRT Wide Field Camera (WFCAM; \citealt{2007A&A...467..777C}). The survey has a resolution of $\sim1\arcsec$, and $5\sigma$ magnitude limits are 19.77, 19.00, and 18.05 in J, H, and K band, respectively \citep{2007astro.ph..3037W}. In our analysis, we use sources with an error less than 0.2~mag in all the three bands. This ensures the good photometric quality of the sources. Using deep NIR and MIR data, we identify the extra YSOs along with the YSOs identified by \citet{2009ApJ...697..787G} and \citet{2011ApJS..193...25R} in the NAN complex. These YSO candidates are used to quantify the star formation properties such as SFR and SFE within the NAN complex.

\subsection{Molecular line data}
Properties of molecular cloud associated with the NAN complex have been analyzed using J = 1 -- 0 transition lines of $\rm ^{12}CO~(115.271204~Hz)~and~^{13}CO~(110.201353~Hz)$ by \citet{2014AJ....147...46Z}. Details regarding the molecular line observations and data reduction can be found in \citet{2014AJ....147...46Z}. The $\rm ^{12}CO~and ^{13}CO$ maps have velocity resolutions of 0.16 and 0.17~$\rm km~s^{-1}$, respectively. Our analysis uses the same molecular line data to probe the gas properties in the NAN complex.

\begin{figure}
\centering
\includegraphics[scale=0.5]{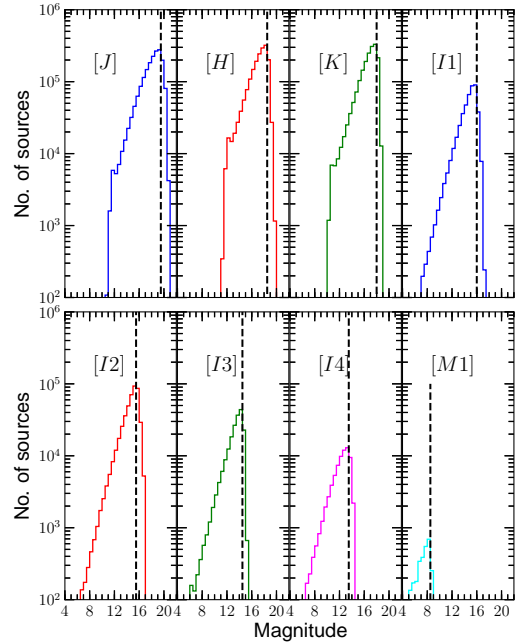}
\caption{Histogram of UKIRT JHK and {\it Spitzer} IRAC, MIPS photometry for all the sources detected within the NAN complex. The dashed line is the turnover point in the individual band, which indicates ~90\% completeness limits for each band.
  }
\label{UKIRT_JHK_hist}
\end{figure} 

\begin{table}
\centering
\caption{Summary of the completeness limits in various bands.}
\label{catalog_num}
\begin{tabular}{ccc}
\\ \hline \hline
Band & Total No. & 90$\%$ Completeness \\
     &  Sources  & Limit (mag) \\
\hline
J  & 1792215 & 19.5 \\
H  & 1792215 & 18.5 \\
K  & 1792215 & 18.0 \\
I1 & 450560  & 16.0 \\
I2 & 450978  & 15.5 \\
I3 & 189723  & 15.0 \\
I4 & 72417   & 14 \\
M1 & 3932    & 8.5 \\
\hline                
\end{tabular}
\end{table}

\section{Analysis and results}

\subsection{Results from NIR and MIR data}
\subsubsection{Completeness of NIR and MIR data }\label{complete_bands}
We analyse the completeness limits of NIR and MIR bands within an area of $2.5^\circ \times 2.5^\circ$ centred at $\rm \alpha = 20:54:59.05$ and $\rm \delta = +44:03:08.58$, using histogram distributions. Histograms in all these bands are shown in Figure \ref{UKIRT_JHK_hist}. The turnover point in the source count is considered as ~90\% completeness of the photometry \citep{{2007ApJ...669..493W},{2013MNRAS.432.3445J},{2015A&A...581A...5S},{2016ApJ...822...49J}}. The completeness limits in all the NIR and MIR bands obtained from the histogram distribution are listed in Table \ref{catalog_num}. This analysis shows the completeness limits of the NIR and MIR photometry within the entire NAN region. However, we caution that the local completeness may vary due to the non-uniform extinction present in the region.

\subsubsection{Additional YSOs in NAN complex}\label{yso_detect}
Selection of candidate YSO towards the NAN region have been carried out by \citet{2009ApJ...697..787G} and \citet{2011ApJS..193...25R} using the photometric data from {\it Spitzer} IRAC and MIPS observations. A total of 2082 YSOs have been retrieved by these authors along with 114 YSOs found in literature which are detected using various techniques. Out of these, 1286 YSOs are from the IRAC and MIPS photometric data (see tables 3 and 4 of \citealt{2011ApJS..193...25R}), 796 YSOs are only from IRAC photometric data, which do not have MIPS counterparts (see table 8 of \citealt{2011ApJS..193...25R}) and remaining 114 are previously suggested YSOs in literatures which are not recovered by IRAC or MIPS (see table 9 of \citealt{2011ApJS..193...25R}). In this study, we attempt to identify the additional YSOs associated with NAN region using the deeper JHK photometric data from UKIRT combining with the photometry in 3.6 and 4.5~$\rm \mu m$ of {\it Spitzer} IRAC. Before identifying the YSOs, we have merged the UKIRT and {\it Spitzer} catalogs, with a cross-matched radius of $1\arcsec$. The merged catalog includes 342586 sources within an area of $2.5^\circ \times 2.5^\circ$ centred at $\rm \alpha = 20:54:59.05$ and $\rm \delta = +44:03:08.58$. 

To identify and classify the YSOs, we used the intrinsic K-3.6 vs. 3.6-4.5 color-color (CC) cut off criteria given in \citet{2009ApJS..184...18G}. To obtain the intrinsic colors of the sources, we made a K-band extinction map of the region based on the (H-K) color excess of the background sources (See \citealt{{2016ApJ...822...49J},{2017ApJ...836...98J}} for details). Here we caution that the systematic errors associated with the adopted extinction laws may lead to an uncertainty in $\rm A_V$ measurements of dense clouds by $\sim$20\% \citep{2016ApJ...822...49J}.

Using the extinction map, we deredden all the sources using the extinction laws described in \citet{2007ApJ...663.1069F}. Following \citet{2009ApJS..184...18G} criteria, we are able to identify 2162 YSOs within the NAN complex. We also have adopted the criteria that all the Class II YSOs must have $\rm [3.6]_0<14.5~mag$, and all protostars must have $\rm [3.6]_0<15~mag$. These magnitude cut-off assures removal of the contaminants such as dim extragalactic sources \citep{2009ApJS..184...18G}.

\begin{figure}
\centering
\includegraphics[scale=0.47]{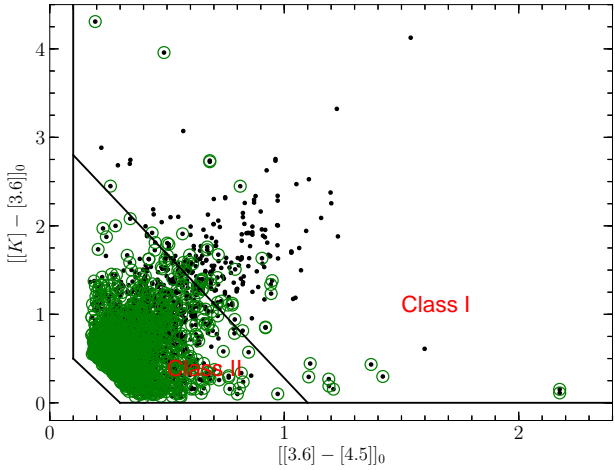}
\caption{$\rm [[3.6]-[4.5]]_0$ vs. $\rm [[K]-[3.6]]_0$ color-color diagram showing the distribution of all the YSOs in NAN complex. The highlighted sources in green color are the additional candidate YSOs identified from this study.}
\label{yso_k36}
\end{figure}

By comparing our YSO list with the list of \citet{2011ApJS..193...25R}, we obtain 1063 YSOs as common sources, and 1099 YSOs are additional detections. The color-color plot of $\rm [[3.6]-[4.5]]_0$ vs. $\rm [[K]-[3.6]]_0$ in Figure \ref{yso_k36} shows the distribution of 2162 YSOs in NAN complex, in which the additional 1099 YSOs identified in this study are highlighted.

\begin{figure}
\centering
\includegraphics[scale=0.47]{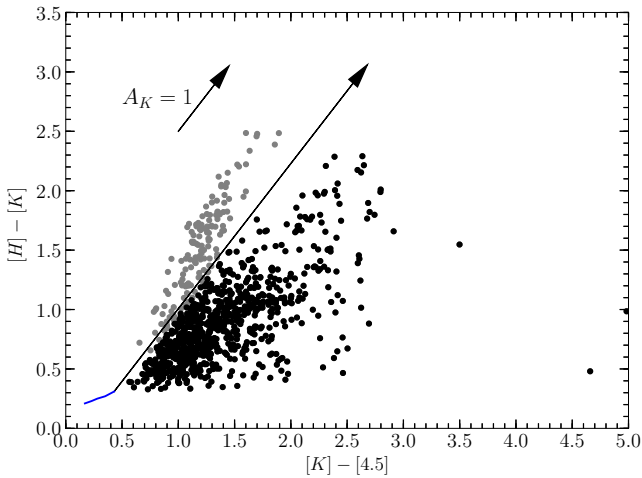}
\caption{Color-color diagram of [K]-[4.5] vs. [H]-[K] to check the YSOs with IR excess. The blue curve shows the locus of late-type M-dwarf stars adopted from \citet{2013ApJS..208....9P}. The long black arrow displays the reddening vector plotted using the extinction laws of \citet{2007ApJ...663.1069F},  starting from the tip of M5 dwarf. All the additional 1099 YSOs are plotted as gray dots, and the 842 IR excess YSOs are overplotted as black dots. A typical length of the reddening vector for $\rm A_K=1~mag$ is also shown on the plot. 
}
\label{yso_hk45}
\end{figure}

In order to remove any possible contaminants in the additional YSO list, we use [K]-[4.5] vs. [H]-[K] CC diagram, which is a well known technique \citep{{2008ApJ...682..445C},{2007ApJ...669..493W},{2014A&A...566A.122S}} to identify YSOs with genuine IR excess emission. 
We plot all the additional 1099 YSOs on the [K]-[4.5] vs. [H]-[K] CC diagram and is shown in Figure \ref{yso_hk45}. The blue curve is the locus of the M-dwarf stars \citep{2013ApJS..208....9P}, in which the color values in H-K and K-W2 (K band minus the WISE band 2) are given. Here we assume the K-W2 is equivalent to K-4.5 since the WISE band 2 and IRAC band 2 have similar central wavelengths. The long arrow is the reddening vector drawn using extinction laws of \citet{2007ApJ...663.1069F}, which starts from the tip of M5 dwarf. The IR excess sources are expected to lie rightward at least 1$\sigma$ away from the reddening vector. Of the additional 1099 YSOs, we find that $\sim$77\% (842 YSOs) lie 1$\sigma$ away to the rightward of the reddening vector. We are unable to retrieve the nature of sources falling leftward of the reddening vector. The majority of these non IR-excess sources are highly reddened with H-K color > 1~mag, which corresponds to to Av > 15~mag, and they show distinct distribution from the majority of the YSOs, and hence they could be the reddened background sources. To be more confident of our detected YSO sample free from any background contaminants, we consider only those YSOs which lie rightward of the reddening vector in Figure \ref{yso_hk45}. Hence, including this conservative constraint into this analysis, we are left with 842 additional YSOs. Figure \ref{yso_hk45} displays the 1099 candidate YSOs identified in this work as small gray dots, and the 842 YSOs with confirmed IR excess are overplotted in black color.

We observe that the Class I YSOs are very less compared to the Class II YSOs, as it is clear from Figure \ref{yso_k36}. Also, we find that these Class I sources are mostly distributed at the location of the {\it Spitzer}, and MIPS identified Class II sources in the H-K vs. 3.6-4.5 CC space, and moreover, we have identified them only using near-infrared photometry (i.e., 2.16 to 4.5~$\rm \mu m$). Hence in this analysis, we consider all the additional YSOs as Class II sources. However, we note that, even if some are true Class I YSOs, it will not affect our results as we are generally interested in the total number of YSOs in this study.

\subsubsection{Final YSO catalog and classification}\label{yso_full}
To make the final YSO list, we used all YSOs (2196) listed in \citet{2011ApJS..193...25R} along with the extra 842 YSOs detected in this study (see section \ref{yso_detect}). Hence after excluding the common sources, the NAN complex is found to be associated with 3038 candidate YSOs. In this section, we obtain the evolutionary stage of the YSOs. Out of the 1286 sources listed in Tables 3 and 4 of \citet{2011ApJS..193...25R}, 273 sources are Class I, 604 are Class II, 112 are Class III, 272 are flat spectrum, and 25 are of unknown type sources. \citet{2011ApJS..193...25R} used [3.6] vs. [3.6]-[24] color-magnitude diagram as the primary mechanism for YSO selection, with adequate care for the removal of contaminants. Table 8 of \citealt{2011ApJS..193...25R}, lists additional 796 YSOs without any evolutionary class, which are not recovered by their YSO classification scheme. These 796 YSOs are a subset of the 1657 YSOs originally detected by \citet{2009ApJ...697..787G} using the four IRAC bands photometry and detection criteria of \citet{2008ApJ...674..336G}. This detection mechanism uses several combinations of color and magnitude cuts using the IRAC bands for identification of YSOs and for the removal of contaminants such as PAH dominated galaxies, AGNs and sources excited with shock emission. To classify the YSOs, \citet{2009ApJ...697..787G} adopt the classification scheme of \citet{1987IAUS..115....1L} and \citet{1994coun.conf..179A}. In this scheme, the YSOs are classified into different classes based on their IRAC spectral index value. Since the evolutionary classes of the 796 YSOs are not mentioned either by \citet{2009ApJ...697..787G} or by \citet{2011ApJS..193...25R}, we have classified the YSOs following the same classification scheme used by \citet{2009ApJ...697..787G}. Out of these 796 YSOs, we find that 58 are Class I, 510 are Class II, 162 are Class III, and 66 are flat-spectrum sources. Finally, we inspected 114 YSOs listed in Table 9 of \citet{2011ApJS..193...25R}, which were detected based on various techniques in the literature. Out of these 114 YSOs, 67 have photometry in all four IRAC bands \citep{2011ApJS..193...25R}. Based on the criteria adopted by \citet{2009ApJ...697..787G}, 3 are retrieved as Class I, 58 are Class II, and 3 are flat spectrum YSOs.  We cross-checked the remaining 47 sources with the YSOs detection technique adopted by us. We retrieve 5 as Class II YSOs, and the rest are assumed to be unknown class.

In summary, from the \citet{2011ApJS..193...25R}, \citet{2009ApJ...697..787G} catalogs, and additional YSOs from our deep JHK and IRAC based data, 3038 YSOs have been identified in NAN complex. Of these, we retrieved 334 as Class I, 341 as flat spectrum, 1964 as Class II, 332 as Class III, and 67 as unknown class YSOs. This is the most complete and updated set of YSOs detected towards the NAN region to date. The full list of 3038 YSOs, along with their evolutionary class, is presented in Table \ref{yso_list}.

In this work, we do not include the Class III sources as these sources are largely incomplete, like the case in Gould Belt (GB) clouds \citep{2015ApJS..220...11D} and represent only 2\% of the total YSO population. To obtain a  full census of Class III objects, an in-depth X-ray survey would be needed \citep{{2010ApJ...719..691F},{2016ApJ...820L..28P}}. However, the missing YSOs may not have a significant effect on the star-forming properties of clumps estimated in this analysis as in the dense regions such as in molecular clumps, majority of the stars might not have reached the Class III phase \citep{{2008ApJ...673L.151G},{2015A&A...581A...5S}}, while their exclusion may underestimate the global SFR and SFE of the entire complex.

\subsubsection{Mass completeness limit of YSOs}
\begin{figure}
\centering
\includegraphics[scale=0.47]{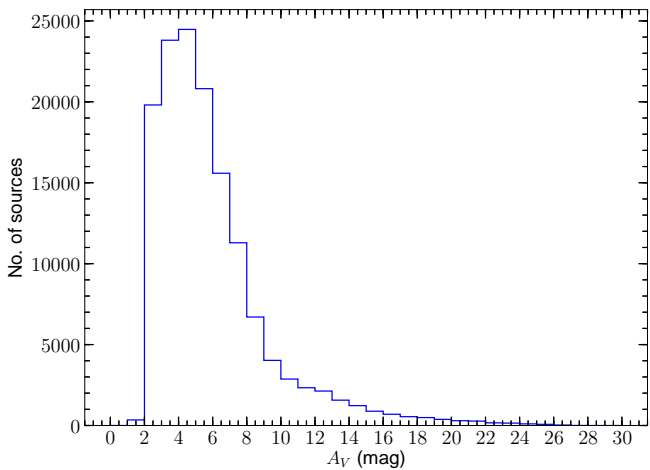}
\caption{Histogram distribution of $\rm A_V$ values towards the NAN complex for the area over which CO molecular line observations have been carried out. This histogram has a bin size of 1~mag.
}
\label{hist_AV}
\end{figure}

As discussed in the previous section, we identified a set of additional YSOs to make the YSO sample more complete. However, the different bands' sensitivities will play a significant role in the identification of the YSOs. In this section, we attempt to determine the mass completeness limit of our detected YSOs following discussions given in \citet{2016ApJ...822...49J} and \citet{2018ApJ...864..154D}. Mass completeness limit varies as a function of extinction of the complex as well as crowding of the region (see Damian et al. 2020 for details). In order to estimate the average value of $\rm A_V$ within the NAN complex, in Figure \ref{hist_AV}, we show the histogram of $\rm A_V$ values of the NAN complex. We find that the NAN complex is associated with a mean $\rm A_V$ value of $\rm \sim 6\pm3.4~mag$.

As discussed in Section \ref{complete_bands}, we see that 90$\%$ completeness limits of UKIRT JHK photometry are 19.5, 18.5, 18.0~mag, respectively. For the {\it Spitzer} 3.6 and 4.5~$\rm \mu m$ photometry, the completeness limits are 16.0 and 15.5~mag, respectively. We use the 4.5~$\rm \mu m$ band to determine the mass completeness limit of YSOs, since this band is shallow compared to other bands. Assuming an extinction range of $\rm A_V = 6- 10~mag$, distance of 800~pc, and considering the pre-main sequence isochrone of 2~Myr \citep{2015A&A...577A..42B}\footnote{The models
described in the paper are available online at http://perso.ens-lyon.fr/isabelle.baraffe/BHAC15dir.
} (details regarding age of 2~Myr is discussed in section \ref{sfr}), the magnitude limit of 4.5~$\rm \mu m$ photometry (i.e., 15.5~mag) corresponds to stellar mass limits of $0.03 - 0.08~\rm M_{\odot}$. This shows that with our deep photometry analysis, we are able to identify the YSOs down to the brown dwarf regime. However, we do not exclude the fact the whole NAN complex suffers from variable extinction. So the local extinction may play a role in the local mass completeness of the YSOs.

\subsubsection{Distribution of YSOs} \label{yso_dist}
\begin{figure*}
\centering
\includegraphics[scale=0.45]{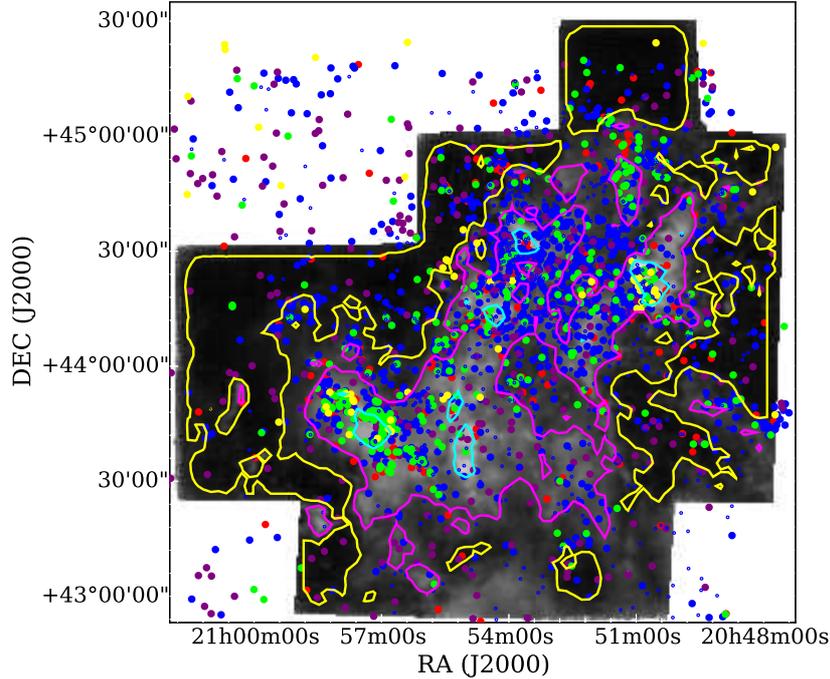}
\caption{Spatial distribution of YSOs on the hydrogen column density image. Red circles are Class I, magenta circles are flat spectrum, blue circles are Class II, purple circles are Class III, and yellow are the unknown YSOs (see Section \ref{yso_full} for details). The contours correspond to hydrogen column density, $\rm N({\rm H_{2}})$ values of $\rm 2.2\times10^{21}~cm^{-2}$ (yellow), $\rm 4.5\times10^{21}~cm^{-2}$ (magenta), and $\rm 8.5\times10^{21}~cm^{-2}$ (cyan), respectively. }
\label{NAN_WISE_yso}
\end{figure*}

The spatial distribution of YSOs on the hydrogen column density image is displayed in Figure \ref{NAN_WISE_yso}. Class I (red), flat spectrum (green), and Class II (blue) are shown in the figure. The YSOs are mostly diagonally distributed from north-west to the south-east. The Class II YSOs dominate towards the north-west region. The south-east shows an overpopulation of Class I and flat-spectrum YSOs. The distribution of YSOs indicates the presence of many sub-clusterings, as suggested in the previous studies \citep{{2002AJ....123.2559C},{2009ApJ...697..787G},{2011ApJS..193...25R}}.

In Figure \ref{NAN_WISE_yso}, the contours represent the distribution of hydrogen column densities ($\rm N({\rm H_{2}})$) of different values. The association of YSOs to different column density levels is seen from this figure. As expected most of the YSOs are located above the $\rm N({\rm H_{2}})$ of $\rm 2.2\times10^{21}~cm^{-2}$. A large fraction of YSOs is distributed above $\rm 4.5\times10^{21}~cm^{-2}$ shows the association of YSOs with dense clumps, suggesting the active star-formation activity within the clumps. Above $\rm 8.5\times10^{21}~cm^{-2}$ very less fraction of total YSOs are located towards the densest part of the cloud. Among the total YSOs $\sim$ 7\%, 5\%, 6\%, 20\% Class I, flat-spectrum, Class II, and Class III YSOs, respectively lie outside the CO mapped region. Thus the majority of projected sources lie inside the cloud area. This also implies that the contamination level of non-YSOs to our Class I to Class II sample may be at the level of 5 to 7\%.

\subsection{Emission from cold dust component} \label{emi_cold}
\subsubsection{Column density from $\rm ^{12}CO$ and $\rm ^{13}CO$} \label{CD_map}
In molecular clouds, star-formation depends on the properties of gas \citep{{2005ApJ...630..250K},{2009ApJ...699..850K},{2007ApJ...668.1064E}}. 
To quantify the properties of the cold dust emission and star-formation, associated with the NAN complex, we have generated the $\rm H_2$ column density map. We estimate the molecular hydrogen column density ($\rm N_{H_2}$) from the $\rm ^{12}CO$ ($\rm I_{^{12}CO}$) and $\rm ^{13}CO$ ($\rm I_{^{13}CO}$) integrated intensity maps. $\rm ^{12}CO$ traces the diffuse part of the molecular cloud, whereas the $\rm ^{13}CO$ traces the dense regions of the molecular cloud. Using both the CO maps, we will be able to get a more accurate estimation of the hydrogen column density. 
At first, we estimate $\rm N_{H_2}$ from $\rm ^{12}CO$ map. The CO-to-$\rm H_2$ conversion factor $\rm X_{CO}$ can be expressed by the following equation \citep{2016MNRAS.456.2406R}.

\begin{equation}
\rm \frac{N_{H_2}(^{12}CO)}{cm^2}~ =~ \rm \frac{X_{CO}}{cm^{-2}~K~km~s^{-1}}~ \rm \frac{I_{^{12}CO}}{K~km~s^{-1}}
\end{equation}

The typical value of the conversion factor $\rm X_{CO}$ in Milky Way is $\rm 2\times10^{20}$ \citep{2013ARA&A..51..207B}. In our analysis, we use the same value to convert $\rm ^{12}CO$ to $\rm H_2$ column density. Further, we estimate $\rm N_{H_2}$ from $\rm ^{13}CO$ using the conversion factor $\rm (^{12}CO/^{13}CO) \times 2.0\times 10^{20}$. We use the $\rm ^{12}CO/^{13}CO$ (mean intensity ratio) to be 3 \citep{2013ARA&A..51..207B}. The final $\rm N_{H_2}$ map is obtained by pixel-wise matching of individual maps generated from the  $\rm ^{12}CO$ and $\rm ^{13}CO$ maps. In the pixel-wise match, we retain the maximum pixel value for the final map. For example, for a particular pixel, if the column density from the $\rm ^{13}CO$ is higher than the $\rm ^{12}CO$, then we retain the higher value. This ensures the contribution from both $\rm ^{12}CO$, which traces the diffuse emission and $\rm ^{13}CO$ map, which traces the dense part of the molecular cloud.
The final column density map is shown in Figure \ref{NAN_H2}. Background value of the column density map is $\rm \sim 1.5\times 10^{21}~cm^{-2}$. We estimate the background level by averaging the $\rm N(H_{2})$ values over a few regions free from diffused emission of the column density map. \citet{2014AJ....147...46Z} find that the lowest contour level of their $\rm ^{12}CO$ map to be $\rm 10~K~ km~s^{-1}$, which corresponds to $\rm \sim 1.8\times 10^{21}~cm^{-2}$ which is similar to our estimated background value of the column density map. The column density map displays several high density clumpy regions. The peak column density value $\rm (\sim 4 \times 10^{23} cm^{-2})$ is located towards the Pelican's Neck region. The average value of column density is $\rm \sim 3.6\times10^{21}~cm^{-2}$.

We have also estimated the column density from $\rm ^{13}CO$ map assuming a local thermodynamic equilibrium (LTE) following the method explained by \citet{1998AJ....116..336N}. As explained above, we combine the column density maps generated from $\rm ^{12}CO$ and $\rm ^{13}CO$ to prepare the final column density map. The average and peak column density values of this column density map are similar to the column density map generated earlier. This shows that the column density map produced assuming LTE is similar to the map generated in the earlier method. We also found that the mass  of the clumps and the entire cloud is in agreement with the estimations of the  first approach within 3 to 8 \% label.

\subsubsection{Associated cold dust clumps} \label{dust_clump}
Figure \ref{NAN_H2} shows the presence of a large number of dense clumps in the region associated with NAN. We use the clump identification algorithm {\it astrodendro} \citep{{2008ApJ...679.1338R},{2009Natur.457...63G}} to detect the dense clumps from the column density map generated in last section. We set a threshold value ($\sim3$ times the background value) for column density to be $\rm 4.5\times10^{21}~cm^{-2}$ and a delta value of $\rm 2\times10^{21}~cm^{-2}$ for the identification of clumps. This threshold value is adopted from the visual inspection, which confirms that most of the emission lie above it, and the optimum choice of delta value helps us to separate the clumps. There are 14 clumps identified within NAN and are marked in Figure \ref{NAN_H2}.

\begin{table*}
\tiny
\centering
\caption{Physical parameters of the cold clumps associated with the NAN region. The peak position of column density, radius, mean column density, total column density, mass, volume density, and gas surface density are listed.
}
\label{clump_properties}
\begin{tabular}{cccccccccc}
\\ \hline
Complex & Clump No. & RA (2000) & DEC (2000) & Radius & Mean $\rm N(\rm H_{2})$ & $\rm \sum N(\rm H_{2}$) & Mass & $\rm n_{H_2}$ & $\rm \Sigma_{gas}^{\dagger}$ \\
& & $(^h~~^m~~~^s)$ & (~$^\circ~~\arcmin~~~\arcsec$) & (pc) & ($\rm \times 10^{21}cm^{-2}$) & ($\rm \times 10^{24}cm^{-2}$) & (M$_{\odot}$) & ($\rm \times 10^{3}cm^{-3}$) & ($\rm M_\odot~ pc^{-2}$)\\
\hline
Pelican's Hat      & 1  & 20:52:36.26 & 44:47:48.47 & 0.9$\pm$0.05 & 5.3 & 10.1 & 306$\pm$2.2  & 1.4$\pm$0.2 & 117.2$\pm$0.9 (0.02$\pm$0.0001) \\
Pelican's Hat      & 2  & 20:51:06.58 & 44:39:40.11 & 1.2$\pm$0.06 & 5.2 & 15.9 & 487$\pm$2.8  & 1.1$\pm$0.2 & 117.1$\pm$0.7 (0.02$\pm$0.0001) \\
Caribbean  Islands & 3  & 20:53:35.74 & 44:32:21.67 & 1.7$\pm$0.08 & 7.2 & 47.2 & 1427$\pm$5.8  & 1.1$\pm$0.2 & 161.4$\pm$0.7 (0.03$\pm$0.0002) \\
Pelican's Beak     & 4  & 20:52:06.41 & 44:22:16.22 & 0.7$\pm$0.03 & 5.4 & 5.3  & 161$\pm$1.6   & 2.0$\pm$0.3 & 120.0$\pm$1.2 (0.03$\pm$0.0001) \\
Pelican's Neck     & 5  & 20:50:50.64 & 44:25:38.18 & 2.0$\pm$0.10 & 8.4 & 80.4 & 2429$\pm$8.0 & 1.0$\pm$0.2 & 187.6$\pm$0.6 (0.04$\pm$0.0002) \\
Caribbean  Islands & 6  & 20:54:59.85 & 44:17:22.84 & 1.6$\pm$0.08 & 8.1 & 45.8 & 1385$\pm$6.0  & 1.3$\pm$0.2 & 181.4$\pm$0.8 (0.04$\pm$0.0001) \\
Pelican's Beak     & 7  & 20:52:09.83 & 44:07:46.51 & 1.5$\pm$0.08 & 7.1 & 38.9 & 1178$\pm$5.0  & 1.1$\pm$0.2 & 159.1$\pm$0.7 (0.03$\pm$0.0001) \\
Pelican's Beak     & 8  & 20:51:42.94 & 43:48:44.10 & 0.5$\pm$0.03 & 5.4 & 3.7  & 111$\pm$1.4   & 2.4$\pm$0.4 & 120.9$\pm$1.5 (0.03$\pm$0.0001) \\
Caribbean  Islands & 9  & 20:55:16.37 & 43:51:22.64 & 1.0$\pm$0.05 & 8.8 & 19.1 & 577$\pm$4.0  & 2.2$\pm$0.3 & 198.3$\pm$1.4 (0.04$\pm$0.0002) \\
Caribbean  Islands & 10 & 20:55:07.92 & 43:33:52.86 & 1.2$\pm$0.06 & 9.6 & 31.7 & 959$\pm$5.3  & 2.0$\pm$0.3 & 213.4$\pm$1.2 (0.04$\pm$0.0002) \\
Gulf of Mexico     & 11 & 20:57:32.10 & 43:47:45.16 & 2.3$\pm$0.11 & 8.1 & 97.5 & 2962$\pm$8.6 & 0.9$\pm$0.1 & 181.2$\pm$0.5 (0.04$\pm$0.0001) \\
Caribbean Sea      & 12 & 20:52:33.10 & 43:41:18.37 & 1.8$\pm$0.10 & 5.7 & 44.1 & 1333$\pm$4.8  & 0.8$\pm$0.1 & 127.9$\pm$0.5 (0.03$\pm$0.0001) \\
                   & 13 & 20:58:25.72 & 43:19:09.70 & 0.7$\pm$0.04 & 6.3 & 7.3  & 219$\pm$2.1  & 2.2$\pm$0.3 & 139.7$\pm$1.3 (0.03$\pm$0.0001) \\
Gulf of Mexico     & 14 & 20:55:59.73 & 43:04:21.90 & 0.8$\pm$0.04 & 5.5 & 8.5  & 256$\pm$2.1  & 1.7$\pm$0.2 & 122.1$\pm$1.0 (0.03$\pm$0.0001) \\
\hline                
Mean   & & & & 1.3 & 6.7 & 32.5 & 985 & 1.5 & 153.4 (0.03) \\
Median & & & & 1.2 & 6.7 & 25.4 & 768 & 1.4 & 149.4 (0.03) \\
Stdev  & & & & 0.5 & 1.5 & 27.8 & 843 & 0.5 & 32.8 (0.007) \\
\hline         
\end{tabular}
\\ $^{\dagger}$ Values of gas surface density ($\rm \Sigma_{gas}$) in parenthesis has unit $\rm gm~cm^{-2}$. 
\end{table*}

\subsubsection{Physical parameters of clumps} \label{phy_clump}
\begin{figure}
\hspace{-0.5cm}
\includegraphics[scale=0.35]{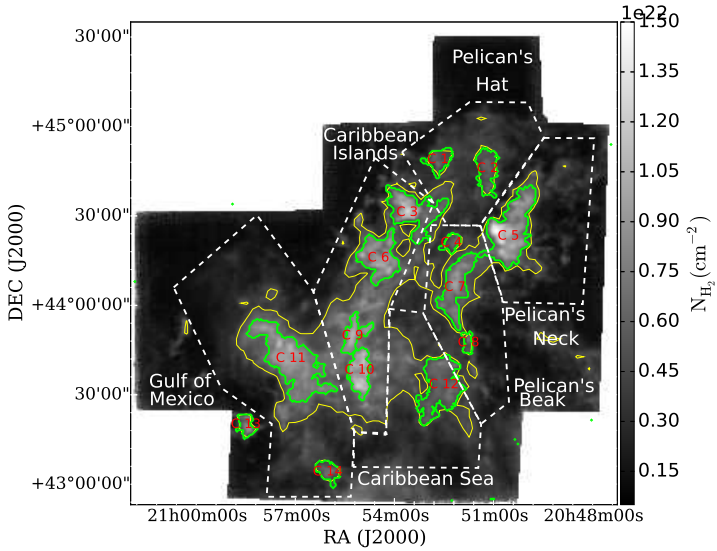}
\caption{Molecular hydrogen column density map generated using the $\rm ^{12}CO$ and $\rm ^{13}CO$ maps integrated over $\rm -20~ to~ +20~km~ s^{-1}$. The green regions represent the boundary of dense clumps along with their numbers. Yellow contours show the threshold of the column density value ($\rm 4.5\times10^{21}~cm^{-2}$) used for detection of clumps in {\it astrodendro} algorithm. Boundaries of the six regions from \citet{2014AJ....147...46Z} overlaid to show the association of dense clumps with the major sub-regions of NAN. }
\label{NAN_H2}
\end{figure} 

We derive several physical parameters such as radius, mass, hydrogen volume number density, and surface density of the star-forming clumps. Using the clump apertures retrieved from {\it astrodendro}, we estimate the physical sizes of the clumps ($\rm r=(A/\pi)^{0.5}$; \citealt{2010ApJ...723L...7K}). Mass of the clumps are calculated using the following equation

\begin{equation}
\rm M_{\rm clump} = \mu_{\rm H_{2}} m_{\rm H} A_{\rm pixel} \Sigma N ({\rm H_{2}}) 
\label{clump_mass1}
\end{equation}

where, $\rm m_{\rm H} $ is the mass of hydrogen, $\rm A_{\rm pixel} $ is the pixel area in $\rm cm ^{2} $, $ \mu_{\rm H_{2}} $ is the mean molecular weight which is assumed to be 2.8 \citep{2008A&A...487..993K}, and $\rm \Sigma N ({\rm H_{2}})$ is the integrated column density. The hydrogen volume number density is derived as, $\rm n_{H_2}= 3 M_{clump} / 4 \pi r^3 \mu_{H_{2}} m_{ H}$, r being the radius. Surface density is derived as, $\rm \Sigma_{gas} = M_{\rm clump} / \pi~r^2$. All the derived physical parameters of the clumps along with the mean, median, and standard deviation are listed in Table \ref{clump_properties}.

We estimate the same physical properties for the entire NAN complex for the area shown in Figure \ref{NAN_H2}. Mass, radius, volume density, and gas surface density of the NAN complex are $\rm (7.5\pm0.1)\times10^4~M_{\odot}$, $\rm 17.2\pm1.0$~pc, $\rm (0.05\pm0.01)\times10^3~cm^{-3}$, and $\rm 81.3\pm0.1~M_\odot~pc^{-2}$ or $\rm 0.02\pm0.0001~gm~ cm^{-2}$, respectively. We compare the physical parameters of the whole NAN region with other nearby Galactic clouds studied by \citet{2014ApJ...782..114E}. Compared to the Galactic clouds, the whole NAN complex is larger in size and massive. The gas surface density of NAN complex matches with the nearby Galactic clouds where the mean and standard deviation value is $\rm 78.6\pm22.2~M_{\odot}pc^{-2}$ \citep{2014ApJ...782..114E}.

\subsection{Measurment of Star-formation rate and efficiency} \label{sfr_sfe}
In this section, we derive the SFR and SFE of the cold dust clumps located towards the NAN region.  These parameters will help us to understand the ongoing star-formation activities within the molecular clumps. Using these parameters we test the various star-formation relations in subsequent sections.

\subsubsection{Star-formation rate} \label{sfr}
We derive the SFR for all the dense clumps using the following relation 

\begin{equation}
\rm SFR = M_{*} / \tau
\end{equation} 

where, $\rm M_{*}$ is the total stellar mass within a molecular clump, and $\rm \tau$ is the average age of the molecular clump. To estimate the SFR, the two important parameters required are age and stellar mass within the star-forming clump.
 \\

For the estimation of age, we follow the discussion made in \citet{2009ApJS..181..321E} and \citet{2010ApJ...723.1019H}. While carrying out a similar analysis, \citet{2009ApJS..181..321E} have assumed an age of 2~Myr, which is an estimate of the time taken to pass the Class II phase. This assumption is made because the study is complete towards the Class II YSOs, or these YSOs are dominant among all the other identified YSOs. As discussed in Section \ref{yso_full}, we have seen a dominance of Class II YSOs ($\sim65\%$ of total YSOs) over all other YSO classes in NAN region. Hence we assume an age of 2~Myr in our analysis. Various uncertainties of SFR estimation in this assumption are discussed by \citet{2010ApJ...723.1019H}. 
\\

The next important parameter in the analyses is the estimation of stellar mass within the dense clumps. Many studies have considered the total mass of YSOs as the stellar mass \citep{{2009ApJS..181..321E},{2010ApJ...723.1019H},{2016ApJ...822...49J}}. 
In our analysis we assume a mean mass of 0.5~$\rm M_\odot$ for YSOs, consistent with the studies of IMF \citep{{2003PASP..115..763C},{2002Sci...295...82K},{2006SerAJ.172...17N}}. Then the total stellar mass is estimated by multiplying the total number of YSOs with 0.5~$\rm M_\odot$.

The derived SFR of the individual star-forming complexes, along with their mean, median, and standard deviation, are listed in Table \ref{clump_sfr}. The uncertainties in the SFR calculations are mainly associated with the YSO counts and we estimate them following the method of \citet{2014ApJ...782..114E}. The SFR values from clump to clump varies in the range of $\rm 0.5\pm0.4 - 78.3\pm4.4~ M_{\odot}~Myr^{-1}$. Clump 11 has maximum SFR, and Clumps 13 and 14 have minimum SFR. The mean of SFR is $\rm \sim 14\pm0.5~M_{\odot}~Myr^{-1}$. The SFR surface density values of the different clumps vary in the range of $\rm 0.2\pm0.1 - 4.8\pm0.5~ M_{\odot}~ yr^{-1}~ kpc^{-2}$. Clump 11 has the maximum SFR surface density, and Clump 14 has the minimum SFR surface density.

For all the 14 clumps, the mean, median and standard deviation of $\rm \Sigma_{SFR}$ are 1.5, 1.0, and 1.2 $\rm M_{\odot}~ yr^{-1}~ kpc^{-2}$, respectively. Also for the complete NAN complex the value of $\rm \Sigma_{SFR}$ is $\rm 0.6\pm0.1 M_{\odot}~ yr^{-1}~ kpc^{-2}$. For the nearby Galactic clouds the mean, median and standard deviation values of $\rm \Sigma_{SFR}$ estimated by \citet{2014ApJ...782..114E} are 0.89, 0.48, and 0.99 $\rm M_{\odot}~ yr^{-1}~ kpc^{-2}$, respectively. Within standard deviation the mean value of $\rm \Sigma_{SFR}$ obtained for the 14 clumps and for the whole NAN region are comparable with the values estimated by \citet{2014ApJ...782..114E}.

\subsubsection{Star-formation efficiency} \label{sfe}
We calculate the star-formation efficiency of the molecular clumps within NAN using the following equation 

\begin{equation}
\rm SFE = \frac{M_{*}}{M_{*}~+~M_{gas}}
\end{equation}

where $\rm M_{*}$ is the total YSO mass within the clump, and $\rm M_{gas}$ is the total mass of clump. The derived SFE of all clumps are listed in Table \ref{clump_sfr}. SFE of all clumps are in the range of $\rm 0.4\pm0.003 - 5.0\pm0.003\%$, and for the whole NAN complex, SFE obtained $\rm 1.6\pm0.004\%$. The mean, median, and standard deviation values of SFE estimated for all the 14 clumps are 2.0, 1.5, and 1.6$\%$, respectively. These values of SFE are consistent with the values obtained for other molecular clouds \citep{{2009ApJS..181..321E},{2010ApJ...724..687L},{2013ApJ...763...51F},{2016ApJ...822...49J}}.

To further study the properties of molecular clouds, we calculate the depletion time for the molecular cloud. The depletion time is expressed as 

\begin{equation}
\rm t_{dep} = M_{gas}~/~\dot{M_{*}}
\end{equation}

where $\rm \dot{M_{*}}$ is the star-formation rate. The derived depletion time for the clouds are listed in Table \ref{clump_sfr}. We find that the depletion time varies in the range of $\sim$~38 -- 548~Myr. Clumps 1, 7, 8, 9, 10, 12, 13, and 14 have $\rm t_{dep}$ more than 100~Myr. These clumps have less SFR, taking a longer time to deplete the gas. For the remaining clumps, $\rm t_{dep}$ are in the range of $\sim$~38 -- 76~Myr. 
The mean, median, and standard deviation of $\rm t_{dep}$ estimated over all the 14 clumps are 218, 132, and 182~Myr, respectively, and for the entire NAN complex, $\rm t_{dep}$ is $\rm 119.6\pm2.4~Myr$. Within the standard deviation, our values agree with the depletion time obtained for the nearby Galactic clouds by \citet{2014ApJ...782..114E}. Their mean, median, and standard deviation of $\rm t_{dep}$ are 201, 106, and 240~Myr, respectively.

\subsubsection{Speed of star formation in clumps}\label{speed_sfr}
The amount of matter that converts into stars over a free-fall time has been addressed by \citet{2007ApJ...654..304K}. They have reported that the conversion of matter into stars follows a slow process, converting only $\sim1\%$ of gas to stars over a free-fall time. The star-formation rate per free-fall time is defined as the fraction of mass of objects converted into stars in a free-fall time at the object's peak density \citep{2005ApJ...630..250K}. We have derived the star-formation rate per free-fall time ($\rm SFR_{ff}$) of the 14 dust clumps of the NAN complex using the relation given below.

\begin{equation}
\rm SFR_{ff} = \frac{\dot{M_{*}} t_{ff}}{M_{gas}} = \frac{t_{ff}}{t_{dep}}
\end{equation}

For each clump, we have derived $\rm t_{ff}$ (explained in Section \ref{vol_law}), $\rm t_{dep}$, and $\rm SFR_{ff}$ are listed in Table \ref{clump_sfr}. The $\rm SFR_{ff}$ of all clumps varies in a range of $0.004 - 0.029$, and for the whole NAN complex the value of $\rm SFR_{ff}$ is 0.0003. Mean, median, and the standard deviation of $\rm SFR_{ff}$ derived for all the 14 clumps are 0.009, 0.006, and 0.008, respectively. The mean $\rm SFR_{ff}$ of clumps within NAN complex are comparable to the values obtained by \citet{2007ApJ...654..304K} and the values obtained by the theoretical predictions by \citet{2005ApJ...630..250K}.

\citet{2014ApJ...782..114E} derived the mean, median, and standard deviation of $\rm SFR_{ff}$ for the nearby star-forming regions to be 0.018, 0.016, and 0.013, respectively. The mean value of $\rm SFR_{ff}$ for the 14 clumps obtained by us and the mean value of nearby star-forming regions obtained by \citet{2014ApJ...782..114E} agree within standard deviation. This suggests that the clumps within the NAN complex and the regions studied by \citet{2014ApJ...782..114E} are equally efficient in converting matter into stars over their free-fall time.

\citet{2019ARA&A..57..227K} reviewed the values of $\rm SFR_{ff}$ estimated based on different studies. They reported that the value of $\rm SFR_{ff}$ estimated from YSO counting and HCN analysis is $\approx0.01$, with a study-to-study dispersion of $\sim$0.3 dex and dispersion of 0.3-0.5~dex within a single study, which is summarized in their figure 10. Our derived median value of $\rm SFR_{ff}$ is 0.006 or -2.2~dex, which is based on YSO counting, matches with other analysis within a dispersion of $\sim$0.3 dex.

\begin{table*}
\tiny
\centering
\caption{YSOs associated with all the cold clumps, densities, star formation rate and efficiency.}
\label{clump_sfr}
\begin{tabular}{cccccccccccc}
\\ \hline
Complex & Clump No. & N (YSOs) & N/Area & SFR & SFR/Area & SFE & $\rm t_{dep}$ & $\rm t_{ff}$ & $\rm t_{cross}$ & $\rm SFR_{ff}$ & $\rm \Sigma_{gas}/t_{multi-ff}$\\
& & & ($\rm pc^{-2}$) & ($\rm M_{\odot}~ Myr^{-1}$) & ($\rm M_{\odot}~ yr^{-1}$ & (\%) & (Myr)& (Myr) & (Myr) & &($\rm M_{\odot}~ yr^{-1}$ \\
& & & & & $\rm kpc^{-2}$) & & & & & & $\rm kpc^{-2}$) \\
\hline
Pelican's Hat      & 1  & 3   & 1.2  & 0.8$\pm$0.4  & 0.3$\pm$0.1   & 0.5$\pm$0.003(0.005) & 408.3$\pm$235.8 & 0.8$\pm$0.06  & 9.0$\pm$2.0  & 0.002 & 168.7$\pm$12.8 \\ 
Pelican's Hat      & 2  & 40  & 9.6 & 10.0$\pm$1.6 & 2.4$\pm$0.5   & 4.0$\pm$0.006(0.040) & 48.7$\pm$7.7  & 0.9 $\pm$0.07 & 19.8$\pm$8.4 & 0.018 & 150.2$\pm$11.6 \\
Caribbean  Islands & 3  & 128 & 14.5 & 32.0$\pm$2.8 & 3.6$\pm$0.5   & 4.3$\pm$0.004(0.043) & 44.6$\pm$4.0  & 1.0 $\pm$0.07 & 22.0$\pm$10.0 & 0.022 & 201.2$\pm$15.5 \\
Pelican's Beak     & 4  & 12  & 8.9 & 3.0$\pm$0.9 & 2.2$\pm$0.7   & 3.6$\pm$0.010(0.036) & 53.6$\pm$15.5  & 0.7 $\pm$0.05 & 7.0$\pm$1.5  & 0.013 & 206.4$\pm$15.7 \\
Pelican's Neck     & 5  & 127 & 9.8 & 31.8$\pm$2.8 & 2.5$\pm$0.3   & 2.6$\pm$0.002(0.026) & 76.5$\pm$6.7  & 1.0 $\pm$0.07 & 37.0$\pm$18.6 & 0.013 & 229.5$\pm$17.2 \\
Caribbean  Islands & 6  & 75  & 9.8 & 18.8$\pm$2.2 & 2.5$\pm$0.4   & 2.6$\pm$0.003(0.026) & 73.9$\pm$8.5  & 0.9 $\pm$0.07 & 20.8$\pm$6.0 & 0.012 & 247.8$\pm$18.8 \\
Pelican's Beak     & 7  & 41  & 5.5 & 10.3$\pm$1.6 & 1.4$\pm$0.3   & 1.7$\pm$0.003(0.017) & 115.0$\pm$18.0  & 0.9$\pm$0.07 & 16.0$\pm$7.5 & 0.008 & 205.9$\pm$15.6 \\
Pelican's Beak     & 8  & 3   & 3.3  & 0.8$\pm$0.4  & 0.8$\pm$0.5   & 1.3$\pm$0.007(0.013) & 148.4$\pm$85.7  & 0.6$\pm$0.05  & 6.0$\pm$1.8  & 0.004 & 229.3$\pm$17.8 \\
Caribbean  Islands & 9  & 8  & 2.8  & 2.0$\pm$0.7  & 0.7$\pm$0.3   & 0.7$\pm$0.002(0.007) & 288.6$\pm$102.1 & 0.7$\pm$0.05  & 14.7$\pm$3.5 & 0.002 & 361.2$\pm$27.7 \\
Caribbean  Islands & 10 & 7  & 1.2  & 1.8$\pm$0.7  & 0.4$\pm$0.2   & 0.4$\pm$0.001(0.004) & 548.2$\pm$207.2 & 0.7$\pm$0.05  & 15.4$\pm$3.8 & 0.001 & 361.8$\pm$27.3 \\
Gulf of Mexico     & 11 & 313 & 19.1 & 78.3$\pm$4.4 & 4.8$\pm$0.5   & 5.0$\pm$0.003(0.050) & 37.9$\pm$2.1   & 1.1$\pm$0.08  & 47.1$\pm$17.8 & 0.029 & 204.4$\pm$15.6 \\
Caribbean Sea      & 12 & 21  & 2.0  & 5.3$\pm$1.1  & 0.5$\pm$0.1   & 0.8$\pm$0.001(0.008) & 253.9$\pm$55.4  & 1.1$\pm$0.08  & 18.9$\pm$5.7 & 0.004 & 135.7$\pm$10.0 \\
                   & 13 & 2   & 1.3  & 0.5$\pm$0.4  & 0.3$\pm$0.2   & 0.5$\pm$0.003(0.005) & 438.7$\pm$312.2  & 0.7$\pm$0.05  & 7.5$\pm$2.7  & 0.002 & 249.2$\pm$18.8 \\
Gulf of Mexico     & 14 & 2   & 1.0  & 0.5$\pm$0.4  & 0.2$\pm$0.1   & 0.4$\pm$0.003(0.004) & 512.2$\pm$362.2 & 0.8$\pm$0.06  & 7.7$\pm$1.7  & 0.002 & 189.1$\pm$14.2 \\
\hline                
Mean   & & 56 & 6.4 & 14.0 & 1.5 & 2.0(0.020) & 218 & 1.0 & 18.0 & 0.009 & 224.3 \\
Median & & 17 & 4.4 & 4.2  & 1.0 & 1.5(0.015) & 132 & 1.0 & 16.0 & 0.006 & 206.2 \\	
Stdev  & & 83 &	5.5 & 21.0 & 1.2 & 1.6(0.016) & 182 & 0.2 & 11.4 & 0.008 & 64.4 \\
\hline
\end{tabular}
We have not listed the error in $\rm SFR_{ff}$, because the values are very small of the order of 5th decimal places. Values of SFE in parenthesis represent their fractional values. 
\end{table*}

\subsection{Test of star-formation relations}
\begin{figure}
\centering
\includegraphics[scale=0.47]{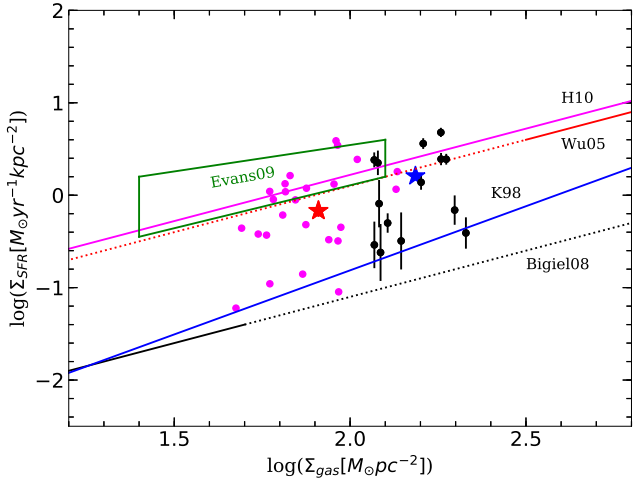}
\caption{The plot between $\rm \Sigma_{SFR}$ and $\rm \Sigma_{gas}$, where the blue solid line shows the relation taken from \citet{1998ApJ...498..541K}. The black line is the relation adopted from \citet{2008AJ....136.2846B}. The solid part of the line is from their study, and the dotted part is extrapolated over higher densities. The red line is from \citet{2005ApJ...635L.173W}, which gives the relation for dense gas from HCN observations. The solid part of this line is from their study, and the dotted part is an extrapolation to lower surface densities. The black filled circles show the values within the 14 clumps from our study. The red and blue stars represent the values obtained over the entire NAN complex and the average value of all the clumps. We also plot the relation (magenta line) between $\rm \Sigma_{SFR}$ and $\rm \Sigma_{gas}$ obtained by \citet{2010ApJ...723.1019H}. The green closed box is for the sources of \citet{2009ApJS..181..321E} taken from their Figure 4, and magenta dots are for the sources of \citet{2014ApJ...782..114E}.
}
\label{NAN_KS}
\end{figure}

\subsubsection{Test of Kennicutt--Schmidt relation} \label{sfr_ks}
The derived value of SFR surface density for the individual molecular clumps towards the NAN complex can be compared with other Galactic star-forming regions and also with other galaxies. We calculate the theoretical value of SFR surface density for all clumps by using the Kennicutt-Schmidt relation, which has been given for other galaxies \citep{1998ApJ...498..541K} as,

\begin{equation}
\begin{split}
\rm \Sigma_{SFR}~ (\rm M_{\odot}~ Myr^{-1}~ pc^{-2})= \\ 
&\hspace{-2cm} \rm ~ (2.5 \pm 0.7)~ \times~ 10^{-4} (\Sigma_{gas} / 1~M_\odot~pc^{-2})^{1.4\pm0.15}
\end{split}
\label{KS_eqn}
\end{equation} 

where $\rm \Sigma_{SFR}$ is the SFR surface density and  $\rm \Sigma_{gas}$ is the gas surface density. In our analyses the gas surface density values for all the clumps are listed in Table \ref{clump_properties}. The predicted values of $\rm \Sigma_{SFR}$ for clumps using equation \ref{KS_eqn} lie in range of 0.2 -- 0.5~$\rm M_{\odot}~ Myr^{-1}$. The observed values of $\rm \Sigma_{SFR}$ for clumps is higher compared to the predicted value. Considering the complete NAN complex, and its $\rm \Sigma_{gas} = 81.3~M_{\odot}~pc^{-2}$, we predict the $\rm \Sigma_{SFR}$ to be 0.12~$\rm M_{\odot}~yr^{-1}~kpc^{-2}$ using the equation \ref{KS_eqn}. The observed $\rm \Sigma_{SFR}$ for the complete NAN complex is 0.7~$\rm M_{\odot}~yr^{-1}~kpc^{-2}$, which is $\sim$6 times higher than the predicted value using the Kennicutt-Schmidt relation.

Figure \ref{NAN_KS} displays the plot between $\rm \Sigma_{SFR}$ and $\rm \Sigma_{gas}$. On the plot, the black filled circles represent our dust clumps, and the red filled star symbol represents the value observed over the entire NAN complex. The average value over all clumps shown as the blue filled star. The relation from \citet{1998ApJ...498..541K} is shown as a blue solid line. The green box on the plot is for the sources of \citet{2009ApJS..181..321E} adopted from their Figure 4. 
In Figure \ref{NAN_KS}, we have plotted SFR-gas relations from \citet{2008AJ....136.2846B} and \citet{2005ApJ...635L.173W}. \citet{2008AJ....136.2846B} have studied the molecular and atomic gas with the sub-kpc resolution of many nearby galaxies and derived the linear relation between the $\rm \Sigma_{SFR}$ and $\rm \Sigma_{gas}$. \citet{2005ApJ...635L.173W} have derived their relation by observing dense gas traced from both Galactic star-forming regions and external galaxies using HCN as a tracer.  \citet{2010ApJ...723.1019H} have also studied many nearby star-forming regions in the solar neighbourhood and derived a relation between the $\rm \Sigma_{SFR}$ and $\rm \Sigma_{gas}$.

In Figure 8, compared to the 14 clumps in NAN, the value of the entire NAN complex lies towards the low $\rm \Sigma_{gas}$ and its position lie among the other Galactic star-forming regions \citep{{2009ApJS..181..321E},{2014ApJ...782..114E}}. Our clumps as well as the sources of \citet{2009ApJS..181..321E} and \citet{2014ApJ...782..114E} lie close to the extrapolated relation of \citet{2005ApJ...635L.173W} and \citet{2010ApJ...723.1019H}. However, all the points are lying above the relation of \citet{1998ApJ...498..541K}. This is likely due to the fact that the relation of \citet{1998ApJ...498..541K} and its coefficient, and exponents are derived by averaging larger regions such as galaxies, where most of the area might be filled with diffused gas without active star formation. This relation might not help to explain the underlying process going on within the star-forming regions of the denser gas of smaller scales. For more details see \citet{2009ApJS..181..321E}.

\subsubsection{Test of Volumetric star-formation relation}\label{vol_law}
\begin{figure}
\centering
\includegraphics[scale=0.47]{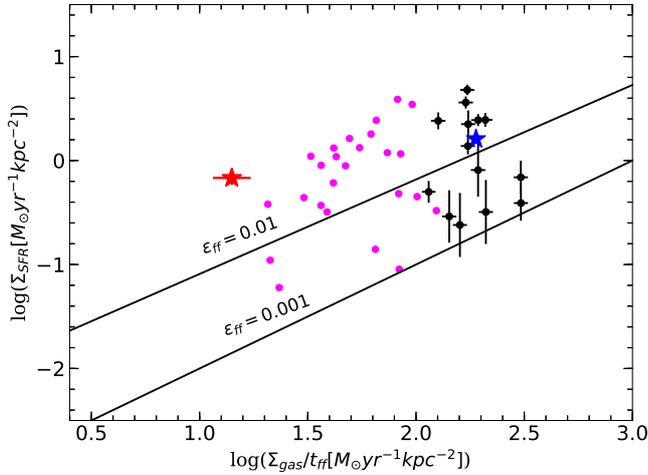}
\caption{This plot shows the variation of $\rm \Sigma_{SFR}$ with $\Sigma_{gas}/t_{ff}$. The two solid lines correspond to the volumetric star formation relation with $\rm \epsilon_{ff}=0.01$ (top) and $\rm \epsilon_{ff}=0.001$ (bottom). The black, magenta dots, and the blue, red stars have same meaning as in Figure \ref{NAN_KS}.}
\label{NAN_sfr_tff}
\end{figure}

In previous section, we check the Kennicut-Schmidt relation. \citet{2012ApJ...745...69K} have argued that the $\rm \Sigma_{SFR}$ is better correlated with $\rm \Sigma_{gas}/t_{ff}$, than $\rm \Sigma_{gas}$ itself, where $\rm t_{ff}$ is the free-fall time scale. This relation between $\rm \Sigma_{gas}/t_{ff}$ and $\rm \Sigma_{SFR}$ is called the volumetric star-formation relation. According to this relation, a fraction of the molecular gas is converted into stars over a free-fall time period. The general form of the relation is expressed as follows \citep{2012ApJ...745...69K}

\begin{equation}
\rm \Sigma_{SFR} = \epsilon_{ff} \frac{\Sigma_{gas}}{t_{ff}}
\end{equation}

where $\rm \epsilon_{ff}$ is a dimensionless measure of the SFR, and it is a constant quantity. \citet{2013MNRAS.436.3167F} defined the quantity $\rm \epsilon_{ff} = \epsilon~\times~SFE$, where $\rm \epsilon_{ff}$ is the local core-to-star efficiency and $\rm \epsilon$ is the fraction of infalling gas, that is accreted by the star.

To test this star-formation relation, we estimate the free-fall time scale of the 14 dust clumps using the following relation 

\begin{equation}
\rm t_{ff} = \left( \frac{3~\pi}{32~ G \mu m_{H} n_{H_2}} \right)^{1/2}
\end{equation}  

where, $\rm \mu = 2.8$, $\rm m_{H}$ is the mass of the hydrogen atom, and $\rm n_{H_2}$ is the volume number density of each clump, and G is the Gravitational constant. For each clump, the volume number density is listed in Table \ref{clump_properties}. Importing these values into above equation, we calculate $\rm t_{ff}$ and are listed in Table \ref{clump_sfr}. Mean, median, and standard deviation of $\rm \Sigma_{gas}/t_{ff}$ for the 14 clumps are 189.2, 174.0, and 54.4~$\rm M_{\odot}~yr^{-1}~kpc^{-2}$, respectively. Value of $\rm \Sigma_{gas}/t_{ff}$ obtained over entire NAN region is $\rm 14.1\pm 2.8~M_{\odot}~yr^{-1}~kpc^{-2}$. 

In Figure \ref{NAN_sfr_tff}, we show the variation of $\rm \Sigma_{SFR}$ with $\rm \Sigma_{gas}/t_{ff}$. In the plot, our clumps are shown as the black dots, the entire NAN region is shown as a red star and the average value of all clumps is displayed as a blue star.  Also, we have plotted the results of \citet{2014ApJ...782..114E} for comparison, which has an average value of $\rm 63.6\pm 32.6~M_{\odot}~yr^{-1}~kpc^{-2}$. The solid lines in the plot correspond to $\rm \epsilon_{ff}=0.01$ (top) and  $\rm \epsilon_{ff}=0.001$ (bottom). All the 14 dust clumps fall above the $\rm \epsilon_{ff}=0.001$, while 10 clumps fall above the $\rm \epsilon_{ff}=0.01$ line. 
Since we have only 14 dust clumps, we have not carried out any fitting to derive the $\rm \epsilon_{ff}$ value. Estimation from this small number will result in a large fitting error. The mean, median, and the standard deviation of $\rm \epsilon_{ff}$ (derived in Section \ref{speed_sfr}) are 0.009, 0.006, and 0.008, respectively. The median value is $\sim 2$ times lower than the value $\rm \epsilon_{ff}=0.01$ derived by \citet{2012ApJ...745...69K}, which is mainly obtained from local to high-redshift galaxies. \citet{2013MNRAS.436.3167F} analysed this relation for many sources and suggested a better fit of this relation over the earlier simple Kennicut-Schmidt relation. However, we see a large scatter among our clumps and also for the sources of \citet{2014ApJ...782..114E}.

\subsubsection{Test of Orbital time model}

\begin{figure}
\centering
\includegraphics[scale=0.47]{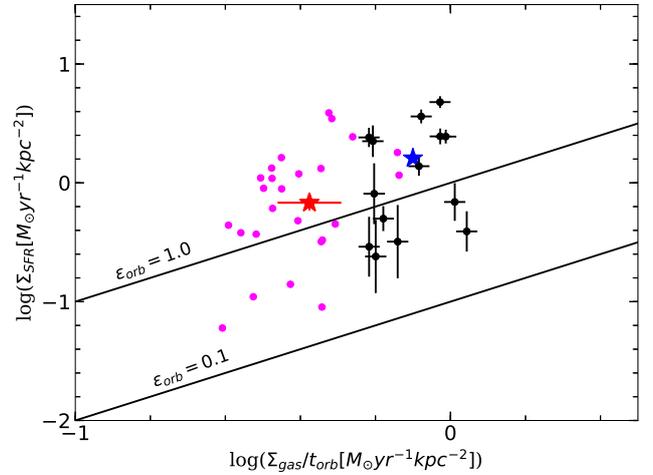}
\caption{This plot shows the variation of $\rm \Sigma_{SFR}$ with $\Sigma_{gas}/t_{orb}$. The two solid lines correspond to the global star formation relation with $\rm \epsilon_{orb}=1.0$ (top) and $\rm \epsilon_{orb}=0.1$ (bottom). The black, magenta dots, and the blue star have same meaning as in Figure \ref{NAN_KS}.}
\label{NAN_sfr_torb}
\end{figure}

\citet{1998ApJ...498..541K} has proposed an empirical star-formation relation, which shows that the SFR of the galaxies depends on the galactic orbital dynamics. This relation can be expressed as follows

\begin{equation}
\rm \Sigma_{SFR} = \epsilon_{orb}\frac{\Sigma_{gas}}{t_{orb}}
\end{equation}

where $\rm \Sigma_{gas}/t_{orb} = \Omega$ is the angular rotation speed, $\rm t_{orb}$ is the orbital rotation speed, and $\epsilon_{orb}$ is the efficiency of star-formation. This relation also called as the $\rm Gas-\Omega$ relation. From the galaxy samples, \citet{1998ApJ...498..541K} has estimated $\rm \epsilon_{orb} = 0.017$, which means that the SFR is $\sim 10\%$ of the available gas mass per orbit. This large scale process, which regulates the formation of stars by affecting the gas over an orbital time period, has been observed towards many galaxies \citep{{2000ApJ...536..173T},{2014ApJ...787...68S}}.

In this analysis, we test this star-formation relation towards the 14 dust clumps of the NAN complex. Whether a large scale effect plays any role on the star-formation process in a Galactic star-forming region can be verified by testing this relation. We calculate the orbital time period of the clumps by using the following relation

\begin{equation}
\rm t_{orb} = \frac{2 \pi R_{G}}{V_{\theta}}
\end{equation}

where, $\rm R_G$ is the Galactic radius, and  $\rm V_{\theta}$ is the azimuthal velocity for a flat rotation curve. We used $\rm V_{\theta} = 254~km~s^{-1}$ from \citet{2009ApJ...700..137R}. Using the relation from \citet{2008ApJ...684.1143X}, we have estimated the Galactic radius towards the NAN region to be 8~kpc. Since all the dust clumps are associated with the NAN complex, so the orbital time period will be same for all, which is estimated to be 193.7$\pm$30.0~Myr. We derived the mean, median, and standard deviation of $\rm \Sigma_{gas}/t_{orb}$ for the clumps and the values are 0.8, 0.8, and 0.2~$\rm M_{\odot}~yr^{-1}~kpc^{-2}$, respectively and for the entire NAN complex the $\rm \Sigma_{gas}/t_{orb}$ is $\rm 0.4\pm0.02~M_{\odot}~yr^{-1}~kpc^{-2}$.

Figure \ref{NAN_sfr_torb} shows the variation of $\rm \Sigma_{SFR}$ with $\rm \Sigma_{gas}/t_{orb}$. The black solid lines on the plot correspond to $\rm \epsilon_{orb}=1.0$ (top) and $\rm \epsilon_{orb}=0.1$ (bottom). All our dust clumps are shown as black dots, and the value obtained over the entire NAN is shown as the red star. The average value obtained overall clumps is shown as the blue star. We have also shown the sources studied by \citet{2014ApJ...782..114E} as magenta dots. For their sources, at first, we estimate the value of $\rm R_G$ and then $\rm t_{orb}$. We used  the Galactic latitudes and longitudes of the sources and the relation of \citet{2008ApJ...684.1143X} to estimate the value of $\rm R_G$. For all the sources of \citet{2014ApJ...782..114E}, we estimate $\rm R_G$ to be $\rm \sim8~kpc$, similar to the value used by \citet{2016A&A...588A..29H}. The values of $\rm t_{orb}$ for sources of \citet{2014ApJ...782..114E} lie within a range of $185 - 204$~Myr. Mean, median, and standard deviation of $\rm t_{orb}$ for sources of \citet{2014ApJ...782..114E} are 193, 191, and 4~Myr, respectively.

In Figure \ref{NAN_sfr_torb}, we see that all our dust clumps distributed around the efficiency of $\rm \epsilon_{orb}=1.0$, which is higher than the value of $\rm \epsilon_{orb}=0.1$ estimated by \citet{1998ApJ...498..541K} for disk galaxies. Similar results are also seen by \citet{2016A&A...588A..29H}, for their ATLASGAL sources. This implies that the large scale processes might play a significant role on the formation of molecular clouds from the ISM \citep{{2016ApJ...823...76K},{2016A&A...588A..29H}}. However, this has no impact on the star-formation processes happening within small-scale regions, like the molecular dust clumps.

\subsubsection{Test of Crossing time model}

\begin{figure}
\centering
\includegraphics[scale=0.47]{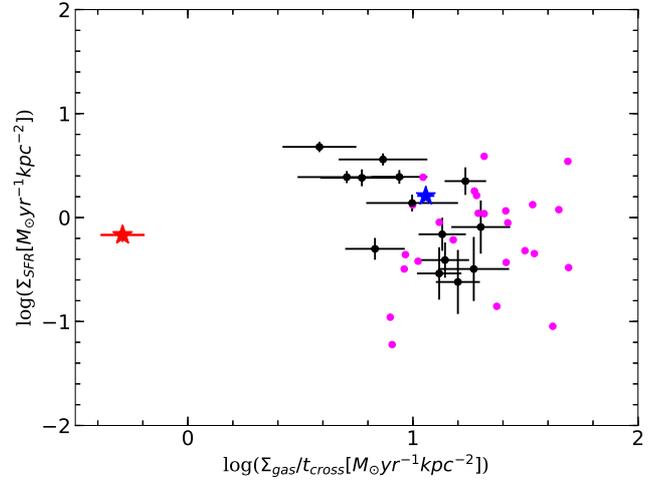}
\caption{This plot shows the variation of $\rm \Sigma_{SFR}$ with $\Sigma_{gas}/t_{cross}$. The black, magenta dots and the blue star have the same meaning as in Figure \ref{NAN_KS}.
}
\label{NAN_sfr_tcross}
\end{figure}

The crossing time scale is defined as the time taken by the disturbance to cross the entire cloud traveling at a turbulent equivalent of mean speed \citep{2014ApJ...782..114E}. We calculate the crossing time for each clumps using the following relation

\begin{equation}
\rm t_{cross} = \frac{2~r_{cl}}{\sigma_v}
\end{equation}

where, $\rm r_{cl}$ is the radius of clump and $\rm \sigma_v$ is the dispersion velocity. Radius of all clumps are listed in Table \ref{clump_properties}. We estimate ${\sigma_v}$ from the line width measured by averaging over the clumps from our $\rm ^{13}CO$ map. The estimated $\rm t_{cross}$ for all the clumps are listed in Table \ref{clump_sfr}. We derived the mean, median, and standard deviation of $\rm \Sigma_{gas}/t_{cross}$ for the clumps as 11.4, 11.5, and 5.1~$\rm M_{\odot}~yr^{-1}~kpc^{-2}$, respectively and for the entire NAN complex the value of $\rm \Sigma_{gas}/t_{cross}$ is $\rm 0.5\pm0.1~M_{\odot}~yr^{-1}~kpc^{-2}$.

The variation of $\rm \Sigma_{SFR}$ with $\rm \Sigma_{gas}/t_{cross}$ is shown in Figure \ref{NAN_sfr_tcross}. The black dots show the clumps' locations, and the value of entire NAN is displayed as the red star. The blue star represents the average value obtained for all the clumps. For comparison, we have also plotted the sources from \cite{2014ApJ...782..114E} as magenta dots on the plot. In this Figure, we do not see any correlation between $\rm \Sigma_{SFR}$ and $\rm \Sigma_{gas}/t_{cross}$.

The relevance of crossing time scale on the star-formation process in molecular clouds is analyzed by \citet{2000ApJ...530..277E}. This study reported that the formation of stars within the molecular cloud begins by condensing the material, and the process gets completed over several crossing times. They also observed that the star formation in large scale structures is slower than in small scale regions. However, in our clumps, we do not find any evidence of the effect of crossing time scale on the star formation process. Similar conclusion have also been made for nearby sources studied by \cite{2014ApJ...782..114E}.

\subsubsection{Test of multi free-fall time model}
The basic idea of the multi free-fall time scale concept is discussed in Section \ref{intro}. The earlier studies \citep{{2012ApJ...745...69K},{2013MNRAS.436.3167F}} have large scatter for relation between $\rm \Sigma_{SFR}$ with $\rm \Sigma_{gas}/t_{ff}$, where $\rm t_{ff}$ is considered as the $\rm t_{single-ff}$  by \citet{2015ApJ...806L..36S}. To reduce the scatter and formulate a more precise relation \citet{2015ApJ...806L..36S} proposed the equation relating $\rm \Sigma_{SFR}$ with $\rm \Sigma_{gas}/t_{multi-ff}$. In this case they assume that the gas densities (PDF) follow the log-normal distribution as the initial condition of star-formation. The log-normal form of PDF is expressed as follows 

\begin{equation}
\rm p(s)ds = \frac{1}{\sqrt{2\pi \sigma^2_s}}~exp \left(-\frac{(s-s_0)^2}{2 \sigma^2_s} \right)ds 
\label{pdf}
\end{equation}

where, $\rm s=ln(\rho / \rho_0)$ is the mean density, $\rm \sigma_s$ is the variance, which related to the mean density as $\rm s_0 = -\sigma^2_s/2$ \citep{1994ApJ...423..681V}. The quantity $\rm \Sigma_{gas}/t_{multi-ff}$ is the maximum gas consumption rate or multi-freefall gas consumption rate (MGCR). The equation of $\rm \Sigma_{gas}/t_{multi-ff}$ related to $\rm \sigma_s$ is given as follows \citep{2015ApJ...806L..36S}

\begin{equation}
\rm \Sigma_{gas}/t_{multi-ff} = \left(\Sigma_{gas}/t_{ff}\right)~ exp \left(\frac{3}{8} \sigma^2_s \right)
\end{equation}

According to \citet{2011ApJ...730...40P} and \citet{2012MNRAS.423.2680M} the logarithmic dispersion has the following expression 

\begin{equation}
\rm \sigma^2_s = ln \left(1 + b^2 M^2 \frac{\beta}{\beta+1} \right)
\label{sigma_s}
\end{equation}

where $\rm M$ is the sonic Mach number, $\rm b$ is the turbulent driving parameter \citep{{2008ApJ...688L..79F},{2010A&A...512A..81F}}, and $\beta$ is the ratio of thermal to magnetic pressure plasma. 

Using the above equations, it is possible to derive the expression for the multi-freefall correction factor \citep{2015ApJ...806L..36S},

\begin{equation}
\rm C_{multi-ff} = \frac{\left(\Sigma_{gas}/t_{multi-ff}\right)}{\left(\Sigma_{gas}/t_{single-ff}\right)} = \left(1 + b^2 M^2 \frac{\beta}{\beta+1} \right)^{3/8}
\label{correction_multi-ff}
\end{equation}

\begin{figure}
\centering
\includegraphics[scale=0.55]{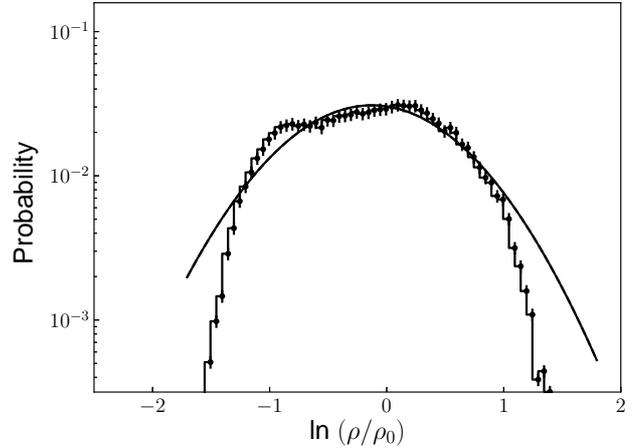}
\caption{Gas density probability density function of the entire NAN complex. This plot is obtained over the full coverage of NAN, as shown in Figure \ref{wise_co}. The solid black line shows the fitting of the log-normal form of PDF given in equation \ref{pdf}. $\rm \rho$ is the density of individual pixel of column density map, $\rm \rho_0$ is average density value obtained by considering the complete NAN region and the probability density points obtained by counting the numbers ($\rm ln(\rho/\rho_0)$) in a bin of 0.05~dex.
 }
\label{NAN_cd_pdf}
\end{figure}

To get an estimate of the logarithmic dispersion we fit the log-normal equation of PDF and fit it to the actual gas density (Figure \ref{NAN_cd_pdf}). In this plot, we estimate $\rm \rho_0$ from the column density map averaging over the entire NAN complex. From the fitting we obtain the value of $\rm \sigma_s = 0.67346\pm0.01695$. In our analysis we use the value of b=0.4 and $\rm \beta \rightarrow \inf$ \citep{{2015ApJ...806L..36S},{2010A&A...513A..67B}}. The use of $\rm \beta \rightarrow \inf$ means that we assume magnetic field to be zero. Importing these parameters into equation \ref{sigma_s}, we derive the value of sonic Mach number $\rm M = 2$, and the multi-freefall correction factor $\rm C_{multi-ff} = 0.6$ for the entire NAN complex. For the 14 clumps associated with NAN, we derive the $\rm \Sigma_{gas}/t_{multi-ff}$ and are listed in Table \ref{clump_sfr}.

\begin{figure}
\centering
\includegraphics[scale=0.47]{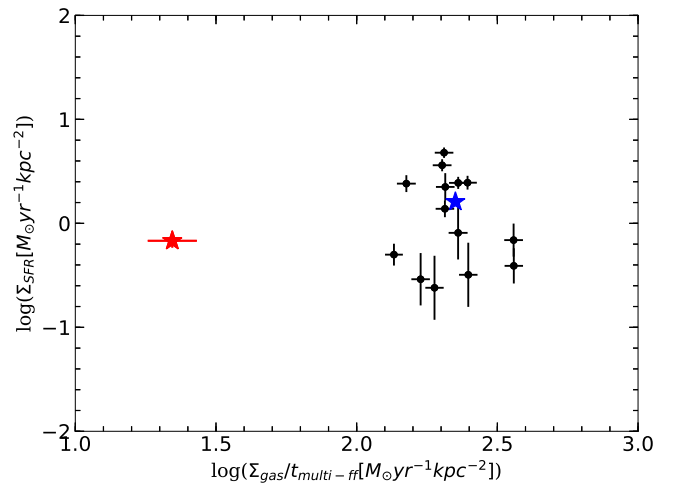}
\caption{This plot shows the variation of $\rm \Sigma_{SFR}$ with $\rm \Sigma_{gas}/t_{multi-ff}$. The black, and the blue star have same meaning as in Figure \ref{NAN_KS}.}
\label{NAN_multi_tff}
\end{figure}

The mean, median, and standard deviation values of $\rm \Sigma_{gas}/t_{multi-ff}$ estimated for the clumps are 224.3, 206.2, and 64.4~$\rm M_{\odot}~yr^{-1}~kpc^{-2}$, respectively and for the entire NAN complex the derived value of $\rm \Sigma_{gas}/t_{multi-ff}$ is $\rm 22.1\pm2.2~M_{\odot}~yr^{-1}~kpc^{-2}$.
In Figure \ref{NAN_multi_tff}, we display the variation of $\rm \Sigma_{SFR}$ with $\rm \Sigma_{gas}/t_{multi-ff}$ for all the 14 dust clumps. From this plot, we could see a similar distribution as we see in Figure \ref{NAN_sfr_tff}, where we show $\rm \Sigma_{SFR}$ with $\rm \Sigma_{gas}/t_{ff}$, which is considered as single free-fall time. This relation received support from recent analysis of few starburst galaxies \citep{{2018MNRAS.477.4380S},{2019MNRAS.487.4305S}}. However, in our case we did not find any significant difference between the variations of $\rm \Sigma_{SFR}$ with respect to $\rm \Sigma_{gas}/t_{multi-ff}$ and $\rm \Sigma_{gas}/t_{ff}$. This  relation also shows a good correlation for regions of large scale compared to smaller star-forming regions.

\subsubsection{Test of star formation rate and dense gas mass}\label{dense_law}

\begin{figure}
\hspace{-0.5cm}
\includegraphics[scale=0.35]{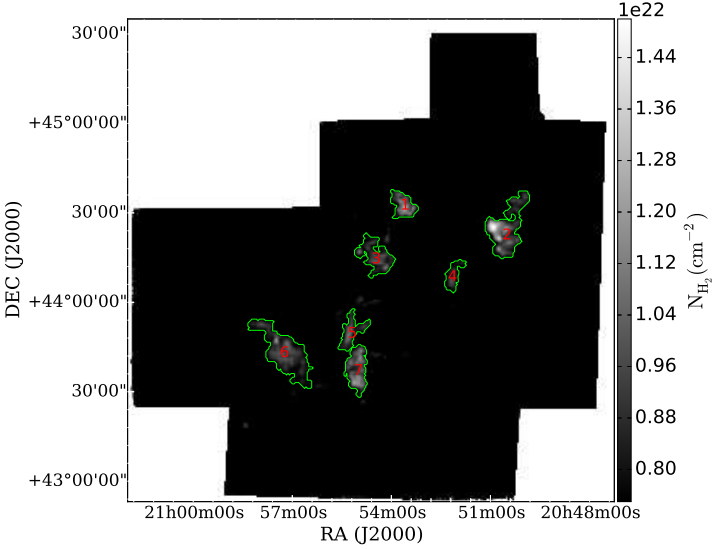}
\caption{Clumps detected above $\rm N ({\rm H_{2}}) = 7.5\times10^{21}~cm^{-2}$ displayed on the column density map. The clump numbers are also given. Scaling of column density map starts with $\rm 7.5\times10^{21}~cm^{-2}$ to highlight only the dense parts of the NAN complex.}
\label{NAN_H2_dense}
\end{figure}

\begin{figure*}
\centering
\includegraphics[scale=0.42]{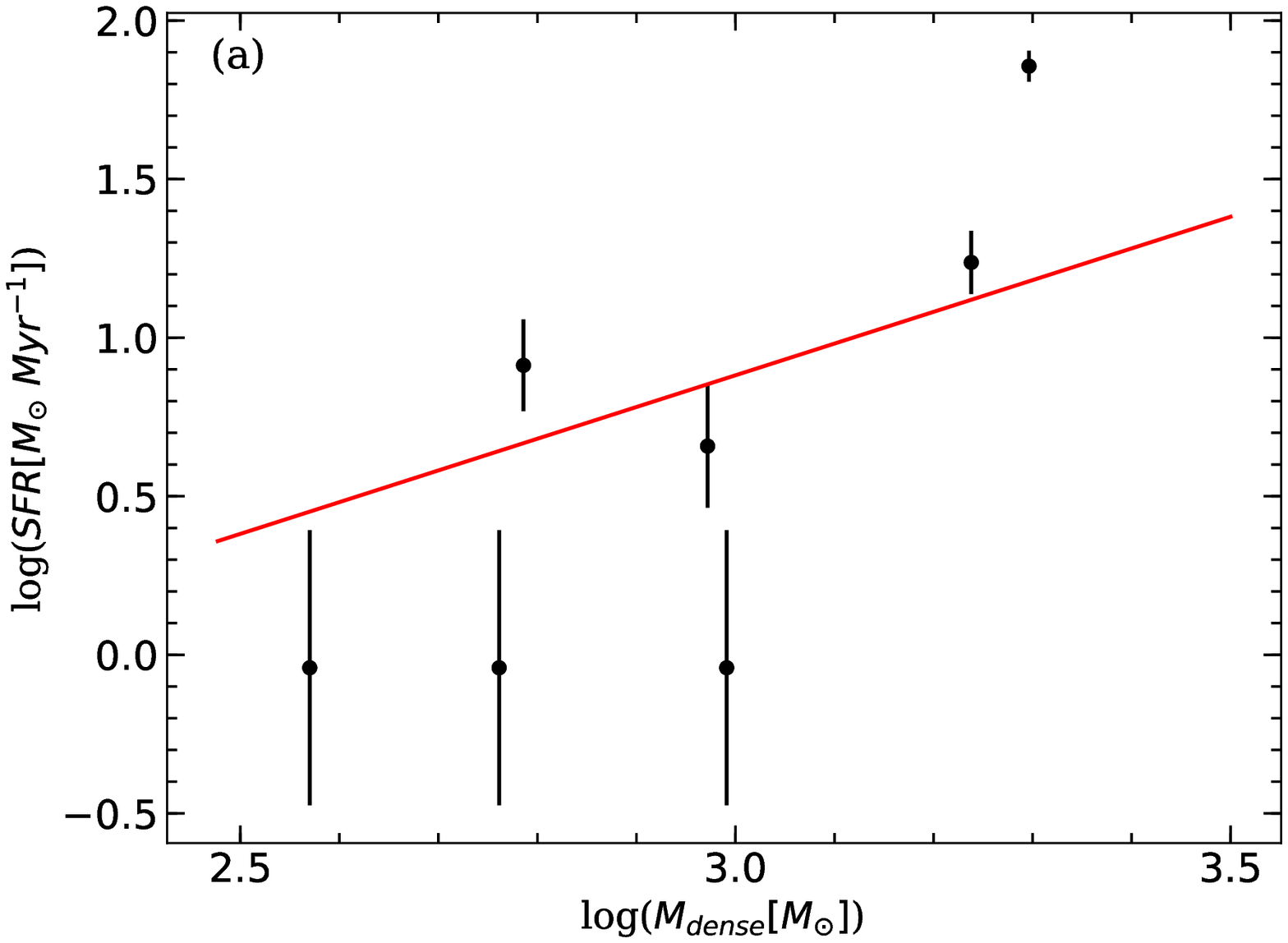}
\includegraphics[scale=0.42]{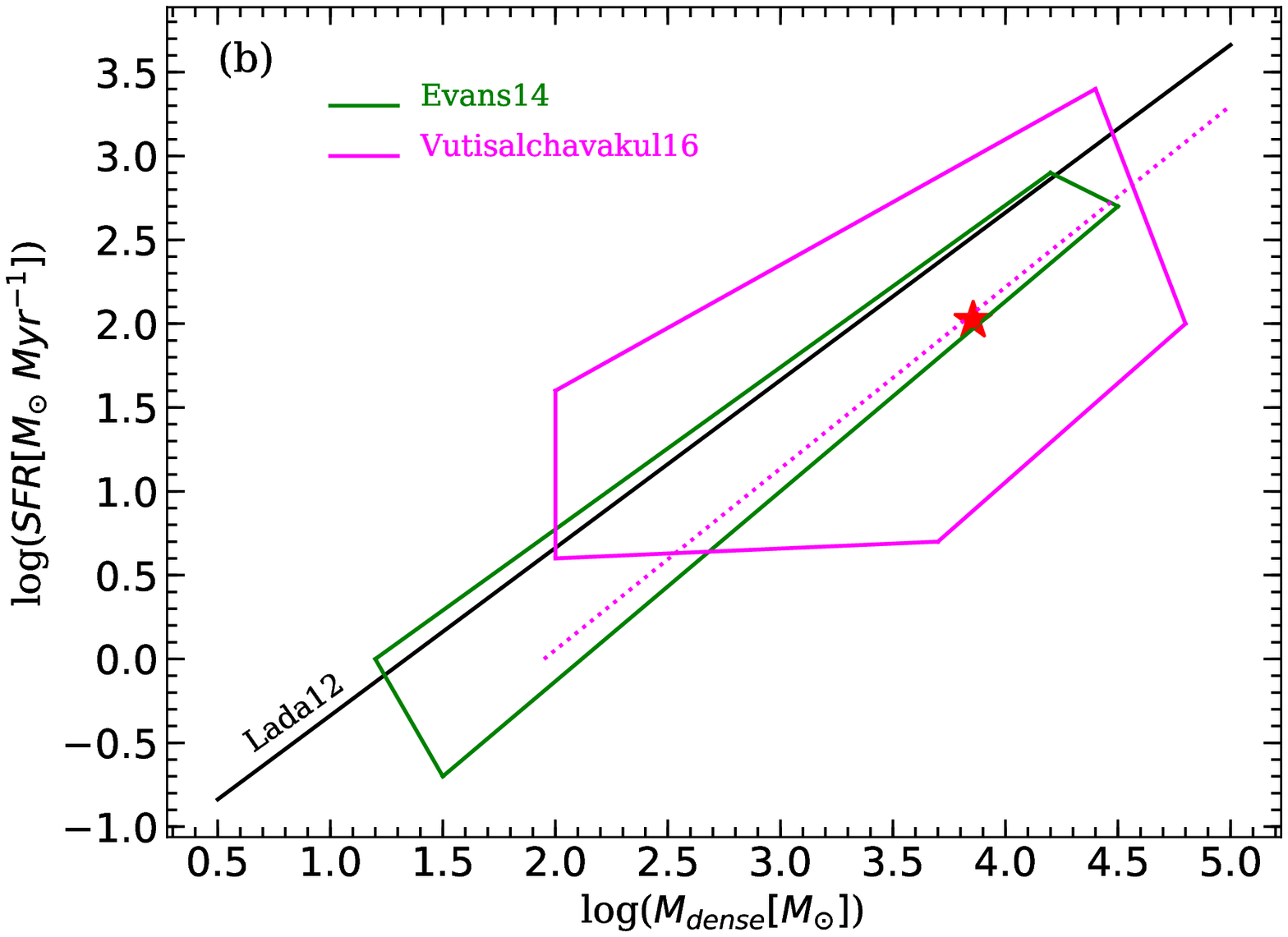}
\caption{(a) Variation of dense mass with SFR for all the clumps detected above $\rm N ({\rm H_{2}}) = 7.5\times10^{21}~cm^{-2}$. The red line is the linear fit considering all the clumps. (b) Variation of dense mass with SFR for the complete NAN complex, which is shown as the red star. Green box demarcates the region covering sources of \citet{2014ApJ...782..114E}, adopted from Figure 8 of their paper. Similarly, the magenta box demarcates the region for sources of \citet{2016ApJ...831...73V} adopted from Figure 4 of their paper. The dotted magenta line is the linear least-square fit adopted from Figure 4 of  \citet{2016ApJ...831...73V}. The prediction of \citet{2012ApJ...745..190L} is plotted as a solid black line. We draw this line using equation 9 of \citet{2014ApJ...782..114E}.
 }
\label{NAN_dense_law}
\end{figure*}

Several studies reported a better and tight correlation between the SFR and the mass of dense gas. \citet{2004ApJ...606..271G} first observed this tight SFR and dense gas relation towards a sample of IR galaxies, and later this has been observed towards the Galactic dense clouds by \citet{2005ApJ...635L.173W}. The idea that the molecular clouds form stars efficiently above a certain threshold has been discussed by many authors \citep{{2008ApJ...680..428G},{2010ApJ...724..687L},{2010ApJ...723.1019H}}. It has been suggested that the SFR is better correlated to the dense mass gas \citep{{2010ApJ...724..687L},{2010ApJ...723.1019H}}. This correlation between SFR and dense gas has been observed in many star-formation studies \citep{{2012ApJ...745..190L},{2014ApJ...782..114E},{2016ApJ...831...73V},{2020ApJ...894..103E}}.

In this work, we test this model towards the dense clumps associated with the NAN complex. We detect these dense clumps from the column density map following a similar procedure mentioned in section \ref{dust_clump}, with a threshold value of $\rm 7.5\times10^{21}cm^{-2}$. We identify seven clumps which lie above this high column density value. Figure \ref{NAN_H2_dense} shows the distribution of these clumps on the column density map. For all these dense clumps, we derive mass, which are listed in Table \ref{dense_clump_sfr}. We call the mass of these clumps as the dense mass. To derive the SFR of the clumps, we count only the Class I YSOs within the clumps and use the age as 0.55~Myr, typically a crossover time of Class I YSOs. The derived SFR values of the clumps are listed in Table \ref{dense_clump_sfr}.

In Figure \ref{NAN_dense_law}(a), we display the relation between the mass of the clumps and their SFR.  The linear fit indicates a relatively poor correlation between SFR and dense gas mass of the clumps. 
In Figure \ref{NAN_dense_law}(b), we plot the dense mass versus SFR of the complete NAN complex (which is equivalent to the sum of the seven clumps) along with the other Galactic star-forming regions found in literature \citep{{2014ApJ...782..114E},{2016ApJ...831...73V}}. For the complete NAN region, mass of dense gas and SFR are estimated to be $\rm 7186\pm14~M_{\odot}$, and $\rm 104.6\pm9.7~M_{\odot}~ Myr^{-1}$, respectively. In Figure \ref{NAN_dense_law}(b), the green box represents the coverage of sources taken from \citet{2014ApJ...782..114E} and the magenta box represents the sources from \citet{2016ApJ...831...73V}. The dotted magenta line displays the linear least-square fit adopted from \citet{2016ApJ...831...73V}. The red star on the plot represents the NAN complex, and the solid black line is the dense gas relation suggested by \citet{2012ApJ...745..190L} drawn using equation 9 of \citet{2014ApJ...782..114E}. From this plot, it is clear that the NAN complex lies among the other Galactic star-forming regions, where the tight relation between SFR and dense gas have already been observed. We agree with the previous analysis of a tight relationship between SFR and dense gas, which is also observed towards the whole NAN region. The dense gas mass is better to predict the SFR of the entire NAN complex rather than for the individual clumps.

\begin{table*}
\tiny
\centering
\caption{Physical parameters of the clumps associated with dense regions. The peak position of column density, radius, mass, gas surface density are listed along with associated YSOs (Class I), densities, and star formation rate.}
\label{dense_clump_sfr}
\begin{tabular}{cccccccccc}
\\ \hline
Clump No. & RA (2000) & DEC (2000) & Radius & Mass & $\rm \Sigma_{gas}^{\dagger}$ & N (YSOs) & N/Area & SFR & SFR/Area \\
& $(^h~~^m~~~^s)$ & (~$^\circ~~\arcmin~~~\arcsec$) & (pc) & (M$_{\odot}$) & ($\rm M_\odot~ pc^{-2}$) & & ($\rm pc^{-2}$) & ($\rm M_{\odot}~ Myr^{-1}$) & ($\rm M_{\odot}~ yr^{-1}~kpc^{-2}$) \\
\hline
1  & 20:53:35.74 & 44:32:21.67  & 0.9$\pm$0.05 & 611$\pm$4  & 213.8$\pm$1.5 (0.04$\pm$0.0002) & 9 & 3.2  & 8.2$\pm$2.7 & 2.9$\pm$1.0 \\
2  & 20:50:50.64 & 44:25:38.18  & 1.6$\pm$0.08 & 1730$\pm$7 & 209.9$\pm$1.0 (0.03$\pm$0.0002) & 19 & 2.3 & 17.3$\pm$3.9 & 2.1$\pm$0.5 \\
3  & 20:54:59.85 & 44:17:22.84  & 1.3$\pm$0.06 & 937$\pm$5  & 192.3$\pm$1.0 (0.04$\pm$0.0003) & 5  & 1.0 & 4.6$\pm$2.0 & 0.9$\pm$0.4 \\
4  & 20:52:09.83 & 44:7:46.51  & 0.8$\pm$0.04  & 372$\pm$3  & 185.4$\pm$1.5 (0.04$\pm$0.0002) & 1  & 0.5 & 1.0$\pm$0.8 & 0.5$\pm$0.4 \\
5  & 20:55:16.37 & 43:51:22.64 & 1.0$\pm$0.05  & 577$\pm$4  & 198.3$\pm$1.4 (0.04$\pm$0.0004) & 1  & 0.3 & 1.0$\pm$0.8 & 0.3$\pm$0.3 \\
6  & 20:57:32.10 & 43:47:45.16 & 1.8$\pm$0.09  & 1979$\pm$7 & 192.1$\pm$0.7 (0.05$\pm$0.0006) & 79 & 7.7 & 71.8$\pm$8.1 & 7.0$\pm$1.0 \\
7  & 20:55:07.92 & 43:33:52.86 & 1.2$\pm$0.06  & 980$\pm$5  & 212.2$\pm$1.2 (0.04$\pm$0.0002) & 1  & 0.2 & 0.9$\pm$0.8  & 0.2$\pm$0.2 \\
\hline                              
Mean   & & & 1.2 & 1027 & 201 (0.04)  & 16  & 2.0  & 15.0 & 2.0 \\
Median & & & 1.2 & 937  & 198 (0.04)  & 5   & 1.0  & 4.6  & 0.9 \\
Stdev  & & & 0.3 & 563  & 11  (0.005) & 26  & 2.5 & 23.8  & 2.2 \\
\hline
\end{tabular}
\\ $^{\dagger}$ Gas surface density ($\rm \Sigma_{gas}$) in parenthesis has unit $\rm gm~cm^{-2}$. 
\end{table*}

\subsubsection{Testing star-formation relations for the entire NAN complex}\label{NAN_full}
\begin{figure*}
\centering
\includegraphics[scale=0.65]{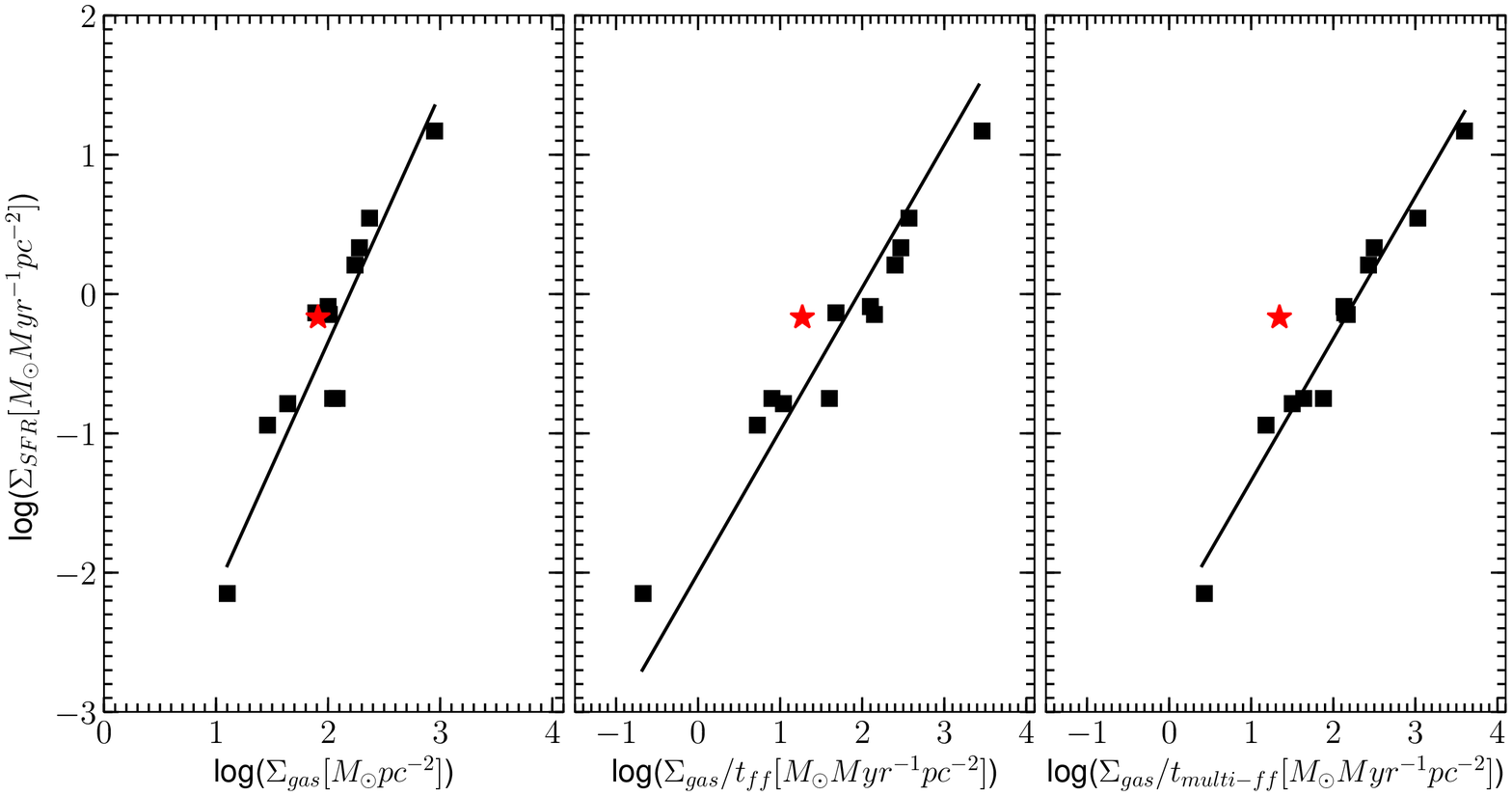}
\caption{This plot shows the variation of $\rm \Sigma_{SFR}$ with $\rm \Sigma_{gas}$,  $\rm \Sigma_{gas}/t_{single-ff}$, and $\rm \Sigma_{gas}/t_{multi-ff}$. The black squares are the sources from \citet{2015ApJ...806L..36S} and the solid lines are from Figure 2 of \citet{2015ApJ...806L..36S}. The red star represents the location of NAN complex.}
\label{NAN_KS_tff_multi_tff}
\end{figure*}

In above sections, we tested several star-formation scaling relations for the dust clumps within the NAN complex. All the star-formation relations show scatter for the 14 dust clumps. Similar scatter was also seen for the nearby star-forming regions analysed by \cite{2014ApJ...782..114E}. However, the ATLASGAL clumps studied by \citet{2016A&A...588A..29H} produced a linear correlation of $\rm \Sigma_{SFR}$ with $\rm \Sigma_{gas}/t{ff}$ and $\rm \Sigma_{gas}/t_{cross}$. Do the relations differ for star-forming regions of different sizes? A more precise relation between $\rm \Sigma_{SFR}$ and $\rm \Sigma_{gas}$ would help to explain the star-formation processes within regions of small size.

In this section, we test the star-formation relations for the complete NAN complex and compare it with the other Galactic regions. \citet{2015ApJ...806L..36S} have compared the three star formation relations, the Kennicutt--Schmidt relation, single free-fall time scale model, and the multiple free-fall time scale model. They study these relations for several Galactic regions and galaxies and is presented in Figure 2 of \citet{2015ApJ...806L..36S}. The different panels of the Figure \ref{NAN_KS_tff_multi_tff} provides a comparison of the three major star-formation relations. Here, we use all the sources analyzed by \citet{2015ApJ...806L..36S} shown as black squares and the NAN complex shown as the red star. The solid black lines are adopted from \citet{2015ApJ...806L..36S}. The location of NAN matches with other Galactic star-forming regions, but with a larger offset from that of \citet{2015ApJ...806L..36S} in the multi free-fall time model. \citet{2015ApJ...806L..36S} obtained the scatter in multi free-fall time scale model to be 1.0, which is less by a factor of 3 -- 4 than the other two star-formation relations implying that this relation is the better to predict SFR. A recent similar analysis by \citet{{2018MNRAS.477.4380S}} and \citet{2019MNRAS.487.4305S} also leads to a similar conclusion. This is because the multi free-fall time scale model accounts for the physical effects such as turbulence and magnetic field along with the effect of gravity. These physical factors play an important role in the formation of stars \citep{{2011ApJ...730...40P},{2012ApJ...761..156F}}.

\section{Discussion}\label{discussion}
We test several star-formation relations in clumps associated with the NAN complex. Here we attempt to find out which star-formation relation is better to predict the SFR. An easy and efficient way is to measure the spread among the data points and check which relation shows the best result. 
For this purpose, we plot the quantity X versus SFR/X in the logarithmic scale. The quantity X is different for different star-formation relations. Then we check which relation is better performing based on their dispersion.

\begin{figure*}
\centering
\includegraphics[scale=0.7]{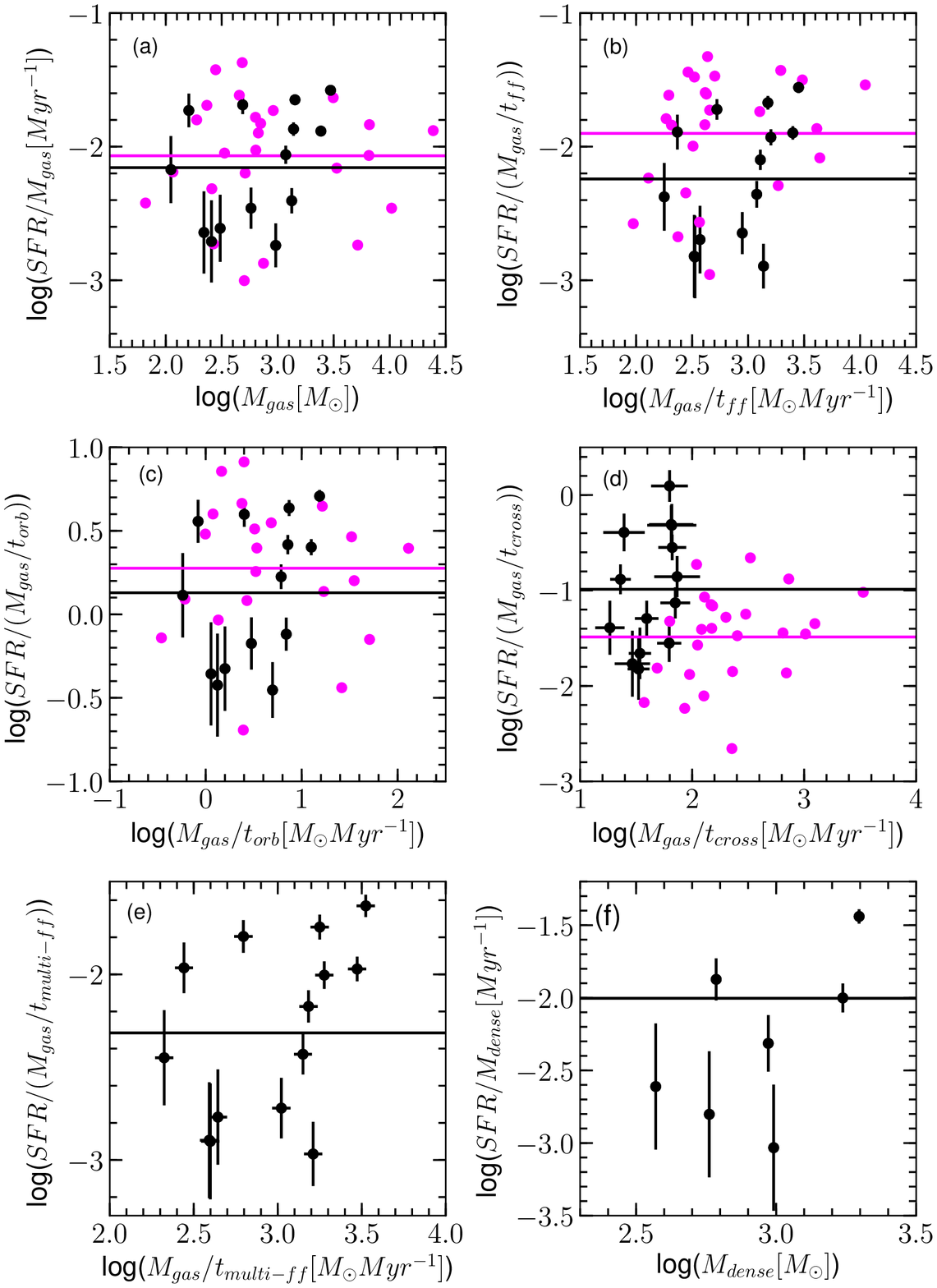}
\caption{Variation of $\rm log(SFR/M_{gas})$ with $\rm log(M_{gas})$ (a), $\rm log(SFR/(M_{gas}/t_{ff}))$ with $\rm log(M_{gas}/t_{ff})$ (b), $\rm log(SFR/(M_{gas}/t_{orb}))$ with $\rm log(M_{gas}/t_{orb})$ (c), $\rm log(SFR/(M_{gas}/t_{cross}))$ with $\rm log(M_{gas}/t_{cross})$ (d), $\rm log(SFR/(M_{gas}/t_{multi-ff}))$ with $\rm log(M_{gas}/t_{multi-ff})$ (e), and $\rm log(SFR/M_{dense})$ with $\rm log(M_{dense})$ (f). On all the plots, the black dots are for clumps associated with NAN complex and the magenta dots are for the nearby star-forming regions taken from \citet{2014ApJ...782..114E}. The black and magenta solid lines represent the mean values.}
\label{NAN_all_ratio}
\end{figure*}

For the Kennicutt-Schmidt relation, X is $\rm M_{gas}$, which is mass of each clump. In the case of volumetric star-formation relation, X is $\rm M_{gas}/t_{ff}$. For the orbital time model, crossing time model, and multi free-fall time model, X is $\rm M_{gas}/t_{orb}$, $\rm M_{gas}/t_{cross}$, and $\rm M_{gas}/t_{multi-ff}$, respectively. For the test of SFR versus dense gas mass, X is $\rm M_{dense}$, which is estimated above $\rm N ({\rm H_{2}}) = 7.5\times10^{21}~cm^{-2}$. In Figure \ref{NAN_all_ratio}, we show the variation of X versus SFR/X. We put all our clumps and compare with nearby Galactic star-forming regions \citep{2014ApJ...782..114E}, wherever possible. The estimated mean and standard deviation of the variable SFR/X are listed in Table \ref{sfr_ratio}.

We observe that for all the relations, the standard deviation in SFR/X is $\sim$ 0.5~dex. The crossing time model displays a maximum dispersion of 0.6~dex. We obtained minimum dispersion to be $\sim$ 0.4~dex for the Kennicut-Schmidt relation and the orbital time scale model. However, it is difficult to make any strong conclusion based on this small number of clumps, about which relation is better to predict the SFR in clumps. 
\citet{2016ApJ...831...73V} have also derived the mean and standard deviation for three star-formation relations (Kennicutt-Schmidt relation, volumetric star-formation relation and for SFR versus dense mass). So we compared the estimation made by us for the clumps with the nearby cloud sample of \citet{2014ApJ...782..114E} and Galactic Plane clouds of \citet{2016ApJ...831...73V}. In all these studies the standard deviation is similar and of the order of $\sim$0.5~dex. However, the mean value differs among the analysis and also among the different predictors. The possible reasons could be the non-uniformity in the various methods used for the estimation of SFR and the uncertainties associated with the assumption that all the clumps are at a similar evolutionary stage that have already produced the majority of the stars.

\begin{table*}
\centering
\caption{Mean and standard deviation of all ratios taken with SFR.}
\label{sfr_ratio}
\begin{tabular}{ccccc}
\\ \hline
Variable & \multicolumn{2}{c}{\underline{~~~~NAN Clumps~~~~}} & \multicolumn{2}{c}{\underline{~~~~Evans 14~~~~}} \\
& Mean & SD & Mean & SD  \\
\hline
$\rm log(M_{gas} [Myr^{-1}])$                        & -2.16 & 0.41 & -2.07 & 0.43 \\
$\rm log(M_{gas}/t_{ff} [M_{\odot} Myr^{-1}])$       & -2.24 & 0.46 & -1.90 & 0.44 \\
$\rm log(M_{gas}/t_{orb} [M_{\odot} Myr^{-1}])$      & 0.13  & 0.41 & 0.28  & 0.40 \\
$\rm log(M_{gas}/t_{cross} [M_{\odot} Myr^{-1}])$    & -1.00 & 0.60 & -1.50 & 0.48 \\
$\rm log(M_{gas}/t_{multi-ff} [M_{\odot} Myr^{-1}])$ & -2.32 & 0.46 & --    & --   \\
$\rm log(SFR/M_{dense} [Myr^{-1}])$       			 & -2.29 & 0.52 & --    & --   \\
													 
\hline            
\end{tabular}
\end{table*}

\section{Summary}
In this work, we have analyzed the entire region of NAN complex for an area of $2.5^\circ \times 2.5^\circ$ and test various star-formation scaling relations in the dust clumps. The major outcome of the work are the following. 

\begin{enumerate}
\item We use the deep NIR data of UKIDSS WFCAM and MIR data in 3.6 and 4.5~$\rm \mu m$ bands of {\it Spitzer} IRAC to identify additional 842 YSOs located in the NAN region. Combining this additional YSOs with the previously detected YSOs by \citet{2011ApJS..193...25R} and \citet{2009ApJ...697..787G}, a total of 3038 YSOs are found to be associated with NAN. This is the deepest and most complete YSO list detected towards the NAN region to date. Out of 3038 YSOs, 334 are Class I, 1964 are Class II, 332 are Class III, 341 are Flat-spectrum, and 67 are unknown type YSOs. The YSOs are complete down to a mass range of $\rm 0.03 - 0.08 M_{\odot}$, for the extinction range $\rm A_V$ of 5 -- 10~mag at a distance of 800~pc and for the average age of $\sim$2~Myr.

\item Most of the YSOs display a spatial distribution along the diagonal region extending from north-west to south-east. Class II YSOs seem to be dominating the north-west region.
 
\item Using the $\rm ^{12}CO$ and $\rm ^{13}CO$ molecular line data, we generate the column density map. Using the {\it astrodendro} algorithm, 14 dust clumps have been identified. Physical properties of all the dust clumps are derived.

\item We derive the SFR and SFE of all the dust clumps and compare the $\rm \Sigma_{SFR}$ with $\rm \Sigma_{gas}$ of all the dust clumps. The SFR and SFE of clumps lies in range of $\rm 0.5\pm0.4 - 78.3\pm4.4~M_{\odot}~Myr^{-1}$ and $\rm 0.4\pm0.003 - 5.0\pm0.003 \%$, respectively. The estimated $\rm \Sigma_{SFR}$ are in range of $\rm 0.2\pm0.1 - 4.8\pm0.5~M_{\odot}~yr^{-1}~kpc^{-2}$. 

\item We test the Kennicutt-Schmidt relation for all the dust clumps. All the clumps lie above the Kennicutt-Schmidt relation. This implies that the relation proposed based on the observations towards galaxies is unable to explain the star-formation processes within local clumps. 

\item From volumetric star-formation relation, we see a large scatter among the clumps. All clumps lie above the efficiency level of 0.001. For this relation, we observe a high scatter among the clumps associated with the NAN complex. 

\item We use the orbital time period model to understand the underlying process of star-formation within the clumps. However, this has no impact on the star-formation process happening within the small scale regions, like the molecular dust clumps.

\item We test the crossing time scale model in all the dust clumps. However, in our clumps, we did not find any evidence of the effect of crossing time scale on the star formation process.

\item The multi free-fall time model has also been tested. Similar to other relations, we did not find any evidence of multi free-fall time on the star-formation within the clumps. 

\item Our test of SFR versus the dense gas mass does not reveal a tight SFR gas relation for the clumps associated with dense gas, which could be due to low statistics of our sample. However, for the whole NAN complex, a tight relation is observed.

\item We test the star-formation relations for the entire NAN complex. We conclude that these star-formation relations have less scatter dealing with galaxies or large scale structures. While for small scale regions, they have large scatter. 

\end{enumerate}

\section*{Acknowledgements}
We would like to thank the referee for valuable suggestions that helped to improve the quality of the paper. We thank Luisa M. Rebull for providing the IRAC and MIPS photometry catalog for the entire NAN complex obtained by the {\it Spitzer} Space Telescope. We thank Neal J. Evans II for his useful comments, which improved this work. This research has made use of NASA's Astrophysics Data System (ADS) Abstract Service, and of the SIMBAD database, operated at CDS, Strasbourg, France. This research has made use of the NASA/IPAC Infrared Science Archive (IRSA), which is funded by the National Aeronautics and Space Administration and operated by the California Institute of Technology. This publication makes use of data products from the Wide-field Infrared Survey Explorer, which is a joint project of the University of California, Los Angeles, and the Jet Propulsion Laboratory/California Institute of Technology, funded by the National Aeronautics and Space Administration.UKIRT is owned by the University of Hawaii (UH) and operated by the UH Institute for Astronomy; operations are enabled through the cooperation of the East Asian Observatory. When (some of) the data reported here were acquired, UKIRT was operated by the Joint Astronomy Centre on behalf of the Science and Technology Facilities Council of the U.K.
This research made use of the data from the Milky Way Imaging Scroll Painting (MWISP) project, which is a multi-line survey in $\rm ^{12}CO/^{13}CO/C^{18}O$ along the northern galactic plane carried out with the Delingha 13.7~m telescope of the Purple Mountain Observatory (PMO). JJ acknowledge the DST-SERB, Government of India for the start up research grant (No: SRG/2019/000664) for the financial support. MWISP project is supported by National Key R$\&$D Program of China with grant no. 2017YFA0402700, and the Key Research Program of Frontier Sciences, CAS with grant no. QYZDJ-SSW-SLH047.

\addcontentsline{toc}{section}{Acknowledgements}

\section*{Data availability}
Table of the list of YSOs is provided as online material. The other data underlying this article will be shared on reasonable request to the corresponding author.




\bibliographystyle{mn2e}
\bibliography{refer} 



\begin{appendix}

\section{YSO list} \label{yso_list}

\clearpage
\onecolumn

\small
{\setlength\tabcolsep{3.5pt} 
\begin{longtable}{@{} l *{11}{c} @{}} \\
\label{YSO_table} \\
\caption{ List of all the 3038 YSOs associated with NAN complex. } \\
\toprule
YSO & RA (J2000) & DEC (J2000) & J & H & K & 3.6 & 4.5 & 5.8 & 8.0 & 24.0  & Class \\
 &$(^h~~^m~~~^s)$ & (~$^\circ~~\arcmin~~~\arcsec$)&(mag)&(mag)&(mag)&(mag) &(mag) & (mag)&(mag) &(mag) & \\  
\midrule
\endhead
\midrule
\multicolumn{12}{r}{Continued on next page} \\
\endfoot
\bottomrule
\endlastfoot
1 & 20:46:24.98 & 44:55:45.83 & 0.00$\pm$0.00 & 0.00$\pm$0.00  & 0.00$\pm$0.00  & 0.00$\pm$0.00  & 0.00$\pm$0.00  &  0.00$\pm$0.00  &  0.00$\pm$0.00 &  4.84$\pm$0.04  &  Unknown  \\ 
2 & 20:46:36.80 & 43:44:34.94 & 0.00$\pm$0.00 & 0.00$\pm$0.00  & 0.00$\pm$0.00  & 9.67$\pm$0.05  & 0.00$\pm$0.00  &  9.15$\pm$0.05  &  0.00$\pm$0.00 &  0.00$\pm$0.00  &  Unknown  \\ 
3 & 20:46:41.54 & 43:46:48.56 & 0.00$\pm$0.00 & 0.00$\pm$0.00  & 0.00$\pm$0.00  & 11.64$\pm$0.05  & 0.00$\pm$0.00  &  11.31$\pm$0.06  &  0.00$\pm$0.00 &  0.00$\pm$0.00  &  Unknown  \\ 
4 & 20:46:45.64 & 43:45:11.43 & 0.00$\pm$0.00 & 0.00$\pm$0.00  & 0.00$\pm$0.00  & 8.15$\pm$0.05  & 0.00$\pm$0.00  &  6.53$\pm$0.05  &  0.00$\pm$0.00 &  0.00$\pm$0.00  &  Unknown  \\ 
5 & 20:47:17.51 & 43:47:49.85 & 0.00$\pm$0.00 & 0.00$\pm$0.00  & 0.00$\pm$0.00  & 12.37$\pm$0.05  & 11.99$\pm$0.06  &  11.64$\pm$0.06  &  11.10$\pm$0.06 &  0.00$\pm$0.00  &  ClassII  \\ 
6 & 20:47:21.51 & 44:10:25.14 & 0.00$\pm$0.00 & 0.00$\pm$0.00  & 0.00$\pm$0.00  & 13.04$\pm$0.06  & 0.00$\pm$0.00  &  12.16$\pm$0.06  &  0.00$\pm$0.00 &  7.52$\pm$0.05  &  Flat  \\ 
7 & 20:47:22.36 & 43:50:11.09 & 0.00$\pm$0.00 & 0.00$\pm$0.00  & 0.00$\pm$0.00  & 11.86$\pm$0.05  & 11.65$\pm$0.05  &  11.47$\pm$0.06  &  10.71$\pm$0.06 &  0.00$\pm$0.00  &  ClassII  \\ 
8 & 20:47:23.59 & 43:44:39.57 & 0.00$\pm$0.00 & 0.00$\pm$0.00  & 0.00$\pm$0.00  & 7.61$\pm$0.05  & 7.17$\pm$0.05  &  6.62$\pm$0.05  &  5.46$\pm$0.05 &  0.00$\pm$0.00  &  ClassII  \\ 
9 & 20:47:25.20 & 43:48:32.05 & 0.00$\pm$0.00 & 0.00$\pm$0.00  & 0.00$\pm$0.00  & 10.01$\pm$0.05  & 9.47$\pm$0.05  &  9.03$\pm$0.05  &  8.13$\pm$0.05 &  0.00$\pm$0.00  &  ClassII  \\ 
10 & 20:47:26.01 & 43:49:8.99 & 0.00$\pm$0.00 & 0.00$\pm$0.00  & 0.00$\pm$0.00  & 14.13$\pm$0.06  & 13.84$\pm$0.06  &  13.63$\pm$0.10  &  12.98$\pm$0.11 &  0.00$\pm$0.00  &  ClassII  \\ 
11 & 20:47:27.03 & 43:49:12.71 & 0.00$\pm$0.00 & 0.00$\pm$0.00  & 0.00$\pm$0.00  & 13.78$\pm$0.06  & 13.57$\pm$0.06  &  13.34$\pm$0.09  &  13.01$\pm$0.14 &  0.00$\pm$0.00  &  ClassIII  \\ 
12 & 20:47:27.89 & 43:50:22.71 & 0.00$\pm$0.00 & 0.00$\pm$0.00  & 0.00$\pm$0.00  & 13.61$\pm$0.06  & 13.38$\pm$0.06  &  13.04$\pm$0.07  &  12.62$\pm$0.08 &  0.00$\pm$0.00  &  ClassIII  \\ 
13 & 20:47:28.17 & 44:57:11.49 & 0.00$\pm$0.00 & 0.00$\pm$0.00  & 0.00$\pm$0.00  & 0.00$\pm$0.00  & 0.00$\pm$0.00  &  0.00$\pm$0.00  &  0.00$\pm$0.00 &  7.32$\pm$0.05  &  Unknown  \\ 
14 & 20:47:30.40 & 43:46:12.19 & 0.00$\pm$0.00 & 0.00$\pm$0.00  & 0.00$\pm$0.00  & 11.10$\pm$0.05  & 10.43$\pm$0.05  &  9.94$\pm$0.06  &  9.46$\pm$0.06 &  0.00$\pm$0.00  &  ClassII  \\ 
15 & 20:47:34.19 & 43:47:23.92 & 0.00$\pm$0.00 & 0.00$\pm$0.00  & 0.00$\pm$0.00  & 11.53$\pm$0.06  & 11.14$\pm$0.06  &  10.50$\pm$0.06  &  9.74$\pm$0.06 &  0.00$\pm$0.00  &  ClassII  \\ 
16 & 20:47:35.34 & 43:44:46.04 & 0.00$\pm$0.00 & 0.00$\pm$0.00  & 0.00$\pm$0.00  & 13.83$\pm$0.06  & 13.56$\pm$0.06  &  13.25$\pm$0.08  &  12.47$\pm$0.09 &  0.00$\pm$0.00  &  ClassII  \\ 
17 & 20:47:36.09 & 44:29:29.97 & 0.00$\pm$0.00 & 0.00$\pm$0.00  & 0.00$\pm$0.00  & 10.10$\pm$0.05  & 0.00$\pm$0.00  &  9.88$\pm$0.06  &  0.00$\pm$0.00 &  7.99$\pm$0.05  &  ClassIII  \\ 
18 & 20:47:37.47 & 43:47:24.81 & 0.00$\pm$0.00 & 0.00$\pm$0.00  & 0.00$\pm$0.00  & 6.62$\pm$0.00  & 6.10$\pm$0.00  &  3.97$\pm$0.05  &  3.52$\pm$0.00 &  0.00$\pm$0.00  &  Unknown  \\ 
19 & 20:47:37.62 & 43:43:37.74 & 0.00$\pm$0.00 & 0.00$\pm$0.00  & 0.00$\pm$0.00  & 10.94$\pm$0.05  & 10.47$\pm$0.05  &  10.10$\pm$0.06  &  9.31$\pm$0.06 &  0.00$\pm$0.00  &  ClassII  \\ 
20 & 20:47:38.03 & 43:49:28.11 & 0.00$\pm$0.00 & 0.00$\pm$0.00  & 0.00$\pm$0.00  & 12.09$\pm$0.05  & 11.76$\pm$0.05  &  11.58$\pm$0.06  &  10.59$\pm$0.06 &  0.00$\pm$0.00  &  ClassII  \\ 
\end{longtable}
(This table is available in its entirety in a machine-readable form in the online journal. A portion is shown here for guidance regarding its form and content.)

}

\end{appendix} 
\end{document}